\documentclass[twocolumn]{aastex62}
\usepackage{microtype}
\usepackage{bold-extra}
\usepackage{graphicx}
\usepackage{amsmath}
\usepackage{multirow}
\usepackage{mathtools}
\usepackage{amstext}
\usepackage{comment}
\usepackage[T1]{fontenc}
\usepackage{apjfonts} 
\usepackage{natbib}
\usepackage[normalem]{ulem}
\newcommand{\lya}{\hbox{Ly$\alpha$}}
\newcommand{\vlya}{\hbox{$\Delta$v$_{Ly\alpha}$}}
\newcommand{\ciii}{\hbox{\sc C\,iii]}}
\newcommand{\siiii}{\hbox{Si\,{\sc iii]}}}
\newcommand{\oiii}{\hbox{\sc [O\,iii]}}
\newcommand{\civ}{\hbox{\sc C\,iv}}
\newcommand{\hi}{\hbox{\sc H\,i}}
\newcommand{\hb}{\hbox{\sc H$\beta$}}

\newcommand{\nv}{\hbox{\sc N\,v}}
\newcommand{\heii}{\hbox{He\,{\sc ii}}}
\newcommand{\hst}{{\it HST}~}
\newcommand{\spitzer}{{\it Spitzer}}

\newcommand{\jwst}{{\it JWST}~}

\newcommand{\ang}{$\mbox{\AA}$}
\newcommand{\sref}{\S\ref} 
\newcommand{\gndone}{z7\_GND\_42912}

\newcommand{\cloudy}{{\sc Cloudy~}}
\newcommand{\bpass}{{\sc bpass~}}
\newcommand{\wciii}{\hbox{W$_{\text{\sc C\,iii}]}$}~}
\newcommand{\editone}[1]{{#1}}

\shorttitle{NIR Spectroscopy of a Galaxy During Reionization}
\shortauthors{Hutchison et al.}

\begin{document}

\title{\large \bf Near-Infrared Spectroscopy of Galaxies During Reionization: Measuring \hbox{\ion{C}{3}]} in a Galaxy at $\mathbf{z=7.5}$}

\author[0000-0001-6251-4988]{Taylor A. Hutchison}
\altaffiliation{NSF Graduate Fellow}
\altaffiliation{TAMU Graduate Diversity Fellow}
\email{aibhleog@physics.tamu.edu}

\author[0000-0001-7503-8482]{Casey Papovich}
\affiliation{Department of Physics and Astronomy, Texas A\&M University, College Station, TX, 77843-4242 USA}
\affiliation{George P. and Cynthia Woods Mitchell Institute for Fundamental Physics and Astronomy,\\ Texas A\&M University, College Station, TX, 77843-4242 USA}

\author[0000-0001-8519-1130]{Steven L. Finkelstein}
\affiliation{Department of Astronomy, The University of Texas at Austin, Austin, TX 78712, USA}

\author[0000-0001-5414-5131]{Mark Dickinson}
\affiliation{National Optical Astronomy Observatory, Tucson, AZ 85719, USA}

\author[0000-0003-1187-4240]{Intae Jung}
\affiliation{Department of Astronomy, The University of Texas at Austin, Austin, TX 78712, USA}

\author[0000-0002-0350-4488]{Adi Zitrin}
\affiliation{Department of Physics, Ben-Gurion University of the Negev, Be'er-Sheva 8410501, Israel}

\author{Richard Ellis}
\affiliation{Department of Physics and Astronomy, University College London, Gower Street, London, WC1E 6BT, UK}

\author[0000-0002-9226-5350]{Sangeeta Malhotra}
\affil{Astrophysics Science Division, Goddard Space Flight Center, Greenbelt, MD 20771, USA}
\affil{School of Earth \& Space Exploration, Arizona State University, Tempe, AZ 85287, USA}

\author[0000-0002-1501-454X]{James Rhoads}
\affil{Astrophysics Science Division, Goddard Space Flight Center, Greenbelt, MD 20771, USA}
\affil{School of Earth \& Space Exploration, Arizona State University, Tempe, AZ 85287, USA}

\author[0000-0002-4140-1367]{Guido Roberts-Borsani}
\affiliation{Department of Physics and Astronomy, University College London, Gower Street, London, WC1E 6BT, UK}

\author[0000-0002-8442-3128]{Mimi Song}
\altaffiliation{NASA Postdoctoral Program Fellow}
\affiliation{Astrophysics Science Division, NASA Goddard Space Flight Center, Greenbelt MD, 20771 USA}

\author[0000-0001-8514-7105]{Vithal Tilvi}
\affil{School of Earth \& Space Exploration, Arizona State University, Tempe, AZ 85287, USA}

\begin{abstract}
We present Keck/MOSFIRE $H$-band spectroscopy targeting \ciii\ $\lambda$1907,1909 in a \editone{$z=7.5056$} galaxy previously identified via \lya\ emission. We detect strong line emission at $1.621\pm0.002\,\mu$m with a line flux of  ($2.63\pm0.52$)$\times10^{-18}$~erg~s$^{-1}$~cm$^{-2}$.  We tentatively identify this line as [\ciii~$\lambda$1907, but we are unable to detect \ciii~$\lambda$1909 owing to sky emission at the expected location. This gives a galaxy systemic redshift, \editone{$z_{sys}=7.5032\pm0.0003$}, with a velocity offset to \lya\ of \vlya\ = \editone{$88\pm27$~km~s$^{-1}$}.  The ratio of combined \ciii/\lya\ is 0.30--0.45, one of the highest values measured for any $z>2$ galaxy.  We do not detect \ion{Si}{3}] $\lambda\lambda$1883, 1892, and place an upper limit on \siiii/\ciii\ $<$ 0.35 ($2\sigma$). Comparing our results to photoionization models, the \ciii\ equivalent width (W$_{\text{CIII]}} = 16.23\pm2.32\,$\AA), low \siiii/\ciii\ ratio, and high implied \oiii\ equivalent width (from the \spitzer/IRAC [3.6]--[4.5]$\simeq$0.8~mag color) require sub-Solar metallicities ($Z\simeq0.1-0.2~Z_{\odot}$) and a high ionization parameter, log\,U $\gtrsim -1.5$.  \editone{These  results favor models that produce higher ionization, such as the \bpass models for the photospheres of high-mass stars}, and that include both binary stellar populations and/or an IMF that extends to 300~M$_{\odot}$.  The combined \ciii\ equivalent width and [3.6]--[4.5] color \editone{are more consistent with ionization from young stars than AGN}, however we cannot rule out ionization from a combination of an AGN and young stars.   We make predictions for  {\it James Webb Space Telescope} spectroscopy using these different models, which will ultimately test the nature of the ionizing radiation in this source. 

\end{abstract}


\section{Introduction} \label{sec:intro}

One of the most important unknowns in extragalactic astronomy is how reionization occurred. During the Epoch of Reionization (z $\approx$ 6--10; EoR), neutral Hydrogen (\hi) dominates the intergalactic medium (IGM), attenuating radiation from early stellar populations and masking galaxies from detection \citep[e.g.,][and references therein]{fink16,star16}. Understanding how and when this occurs can reveal whether or not these young galaxies provided the necessary ionizing radiation to completely reionize the IGM by z $\simeq$ 6 \citep[inferred from quasar Ly$\alpha$ forests, e.g.][]{fan02,mort11,vene13,mcgr15}, less than one billion years after the Big Bang. 

The physical properties of galaxies during this epoch are not well understood, as only a very small number have been confirmed spectroscopically, with only the brightest sources detected \citep[e.g.,][]{vanz11,ono12,schen12,shib12,fink13, oesch15,zitr15,robo16,song16,star17,matt17,lars18,jung19}.  Due to the galaxies' distances, they are very faint \citep[$H_{160} \gtrsim$ 25--27 mag, e.g.][]{fink15,bouw15}, with their (rest-frame) UV spectral features pushed out to near-infrared (NIR) wavelengths. By nature of the sensitivity of current ground- and space-based telescopes, sky brightness, and current observing techniques, spectroscopic surveys invariably have a magnitude limit imposed on the sample. 

Lyman-$\alpha$ ($\lambda_0 = 1216\,$\ang; Ly$\alpha$) emission is one of the most common rest-frame UV features used to study galaxies at higher redshifts ($z \gtrsim 2$) as it is observed to be very strong in star-forming galaxies \citep[e.g.,][]{shap03}. 
Spectroscopic studies target \lya-emitting galaxies (LAEs) and Lyman Break galaxies (LBGs), selected from large broad- and narrow-band NIR surveys \citep[e.g.,][]{fink13}, in order to study the evolution of galaxy properties with redshift. Comparing the \lya\ and continuum UV properties of galaxies yields constraints on the \lya\ escape fraction
\citep[e.g.,][]{hayes10},  which has been used to argue for increasing ionization in galaxies. \editone{In addition, constraints on the \lya\ escape fraction have inferred} an increasing \ion{H}{1} neutral fraction in the IGM at $z>6.5$, \editone{where the neutral gas in the IGM is expected to heavily suppress \lya\ emission except in large, ionized bubbles} \citep[e.g.,][]{hayes11,treu13,dijk14,schen14,tilvi14,star16,mason18}.

However, the observed \lya\ emission is heavily dependent upon the spatial distributions of \hi\ as well as the intrinsic characteristics of its emitting galaxy \citep[e.g.,][and references therein]{matt16,matt17,sobr17}, which makes it a useful line for inferring the properties of the circumgalactic medium (CGM) and the interstellar medium (ISM) when compared to nebular lines \citep[e.g.,][]{moll98,stei11,verh15}. However, one of the complications of using \lya\ (referenced above) is the resonant scattering due to \hi\ \citep[e.g.,][and references therein]{dijk14}, often shifting the emission hundreds of km s$^{-1}$ redwards of the galaxy's systemic (or true) redshift.  This effect is pronounced in galactic outflows, where the blue-shifted portion of the \lya\ emission has been absorbed and scattered away from the line of sight, allowing the redshifted (backscattered) \lya\ emission to pass through unattenuated \citep{shap03,dijk14,erb14}.  This redshifting of \lya\ can add uncertainty to any intrinsic property inferred for a galaxy when using just \lya\ emission \citep[e.g.,][]{hayes15}, as well as significantly impact the visibility of \lya. Due to the small number of current significant detections of galaxies during the EoR (see above) determining the systemic redshift of these sources is an active field of study \citep[e.g.,][]{star14,star15a,star15b,star17,ding17,matt17,main17}.

Recent studies targeted alternative nebular emission lines to measure the systemic redshift of high-redshift galaxies, and give insight into the galaxies' kinematics and ionization \citep[e.g.,][]{star14,star17,ding17,mase17,matt17}.  UV lines from metals are the best candidates, with \ciii\ $\lambda\lambda$1907,1909\,\ang\ as the most promising \citep[e.g.,][]{star15a,star15b,jask16,ding17,matt17}. Observations of star-forming galaxies at 1.5 $<$ z $<$ 8 show that \ciii\ is the strongest UV line after \lya\ emission \citep[e.g.,][]{shap03,erb10,main17}. 
In addition, at $z\sim2$ \ciii\ equivalent widths (W$_{\text{CIII]}}$) appear to be larger for lower metallicity galaxies \citep[e.g.,][]{erb14,star14,naka18b}. With both \lya\ and \ciii\ detections, the systemic (\ciii) and attenuated (\lya) redshifts can be compared to shed light on the structure and ionization of the CGM as well as the IGM \citep{du18}. Understanding the current sample of distant galaxies detected in more detail will better constrain models and enable a deeper understanding of what can be expected with next generation of telescopes.

We have begun a spectroscopic study with Keck/MOSFIRE to measure the rest-frame UV emission properties of high-redshift galaxies at $z > 5$, with the intention of understanding the frequency of emission line fluxes (other than \lya) as a function of galaxy property (including apparent magnitude), and to constrain the physical conditions in the galaxy (galaxy metallicities, ionizing source, etc).  This will inform surveys for spectroscopic redshifts and expectations for forthcoming observations with the {\it James Webb Space Telescope} ({\it JWST}). Here, we present the first results from this work, studying the $H$-band spectroscopy of a galaxy at $z > 7$ with a redshift from \lya, targeting the \ciii\ $\lambda\lambda$1907,1909 emission feature.    

The remainder of this work is outlined as follows.  In~\sref{sec:data}, we present the observations, data reduction, and calibration. The spectra and identification of \ciii\ lines are presented in~\sref{sec:result}; we discuss the implications of our results in~\sref{sec:discuss} and use stellar population models paired with photoionization simulations to synthesize the spectral energy distribution (SED) of the galaxy. Finally, we summarize our conclusions in~\sref{sec:summary}. Throughout we use AB magnitudes \citep{oke83} and adopt a cosmology with $\Omega_M=0.3$, $\Omega_{\Lambda}=0.7$, $\Omega_K=0.0$, and $h=0.7$ (where $H_0=100~h$ km s$^{-1}$ Mpc$^{-1}$) consistent with Planck and local measurements \citep{plan16,riess16}.

\section{Data \& Methods} \label{sec:data}
The $z > 7$ galaxies in our sample were all selected using photometry from the GOODS \citep{giav04} and CANDELS \citep{grog11,koek11} imaging data, combined with photometric redshift measurements and selection methods discussed in \citet[hereafter F13]{fink13} and \cite{fink15}.  Two of the galaxies targeted in our observations were previously confirmed to have $z > 7$ from spectroscopic measurements of \lya\ emission \citep[F13;][hereafter J19]{jung19}.   
The galaxy featured in this work, \gndone, was originally spectroscopically-confirmed via \lya\ with  $z_{Ly\alpha} = 7.5078 \pm 0.0004$ by F13.  This was confirmed by \citet[hereafter T16]{tilvi16} using {\it HST} grism spectroscopy.  Recently, J19 published an updated redshift of  \editone{$z_{Ly\alpha} = 7.5056 \pm 0.0007$} using data from $>$16 hr of total $Y$-band MOSFIRE spectroscopy.\footnote{\editone{Note that J19 find evidence the \lya\ line is asymmetric.  They report a \lya\ redshift using a centroid from a skewed Gaussian.  As a result their \lya\ redshift is systematically different compared to a fit to \lya\ using a symmetric Gaussian.  The latter gives a redshift $z_{Ly\alpha} = 7.5072 \pm 0.0003$, consistent with the original value from F13, who also used a symmetric Gaussian.}}

\subsection{Observations \& Data Reduction} \label{subsec:observations}
$H$-band spectroscopic observations of our sample were taken using the Keck/MOSFIRE NIR spectrograph \citep{mcle12} for three nights in 2014 and one night in 2017.  In addition, we make use of data from an independent observation taken with MOSFIRE in 2016. The dates of observations, total exposure time, average seeing (derived from the final reduced 2D spectrum), and instrument setup are shown in Table \ref{table:1}. For all of the data, a standard dither pattern of ABAB was used, with 0\farcs7 width slits for all targets in the masks and standard star frames. In each mask, a reference star was placed on a slit to monitor the seeing and atmospheric variability as well as improve the flux calibration process by providing an absolute calibration (see~\sref{subsec:flux-cal}). Due to significant photometric variability, the 2014 March 15 data and parts of the 2014 March 14 data were left out of our analysis (see Appendix \ref{appsub:phot}). Ar and Ne arcs were taken, although skylines were used instead for wavelength calibration. 

\begin{deluxetable*}{llccccccccc}[!ht]
\tablecaption{\label{table:1}Keck I Observations Using the MOSFIRE NIR Spectrograph.}
\tablecolumns{11}
\tablehead{
\colhead{Observation} & \colhead{PI} & \colhead{Band} & \colhead{Seeing} & \colhead{$t_{exp}$} & \colhead{Slit Width\tablenotemark{a}} & \multicolumn{1}{c}{Dithering} & \colhead{Stepsize} & \multicolumn{3}{c}{Reference Star}  \\ \cline{9-11}
\colhead{Date} & \colhead{} & \colhead{} & \colhead{(arcsec)} & \colhead{{\small (hr)}} & \colhead{(arcsec)} & \colhead{Pattern} & \colhead{(arcsec)} & \colhead{$\alpha$} & \colhead{$\delta$} & \colhead{$H_{160}$} \\
\colhead{(1)} & \colhead{(2)} & \colhead{(3)} & \colhead{(4)} & \colhead{(5)} & \colhead{(6)} & \colhead{(7)} & \colhead{(8)} & \colhead{(9)} & \colhead{(10)} & \colhead{(11)}}
\startdata
2017 Apr 18 & Papovich & $H$ & 0.7 & 3.711 & 0.7 & ABAB & 1.50 & 189.304712 & +62.269859 & 17.29 \\
2016 May 04 & Zitrin\tablenotemark{b} & $H$ & 0.65 & 1.988 & 0.7 & ABAB & 1.25 & 189.105720 & +62.234683 & 15.99 \\
2014 Mar 25 & Malhotra & $H$ & 0.8 & 0.994 & 0.7 & ABAB & 1.25 & 189.287154 & +62.297020 & 17.12 \\ 
2014 Mar 15 & Finkelstein & $H$ & 1.0 & 1.392\tablenotemark{c} & 0.7 & ABAB & 1.25 & 189.287154 & +62.297020 & 17.12 \\
2014 Mar 14 & Finkelstein & $H$ & 0.6 & 1.789\tablenotemark{c} & 0.7 & ABAB & 1.25 & 189.287154 & +62.297020 & 17.12 \\
\enddata 
\tablecomments{(1) UT Date of observation. (2) PI of Keck observing program. (3) MOSFIRE NIR band observed. (4) Average seeing derived from the final reduced images of the star in the mask. (5) Total exposure time of the mask. (6) Slit widths in the mask. (7) Dithering pattern of the instrument. (8) Stepsize of the dithering. (9-11) RA, Dec, and $H_{160}$ magnitude of the Reference Star in the mask.}
\tablenotetext{$a$}{Standard MOSFIRE slit width: 0\farcs7.}
\tablenotetext{$b$}{Data not used in final spectral coadd.}
\tablenotetext{$c$}{Much of the 2014 Mar 14--15 data are taken under poor conditions and unusable.  We include only data from 2014 Mar 14 taken from 08:20--09:35 UTC and no frames from 2014 March 15 (see Appendix \ref{appsub:phot}).}
\end{deluxetable*}

The data were reduced using the MOSFIRE data reduction pipeline (DRP\footnote{\url{http://keck-datareductionpipelines.github.io/MosfireDRP/}}).
The DRP produces background-subtracted, rectified, and flat-fielded two-dimensional (2D) spectra and associated 2D variance for each slit within a given mask. The resulting spectral resolution yields 1.63 \ang~pixel$^{-1}$, with 0\farcs18 pixel$^{-1}$ spatially. 
We visually inspected the reduced 2D spectra at the potential positions of UV emission lines (such as the \ciii\ doublet) for our galaxies. 

\subsection{Optimized 1D Extraction \& Flux Calibration} \label{subsec:flux-cal}

We extracted one-dimensional (1D) spectra for our sources at the position of an observed emission feature or at the expected center of the slit with an extraction box width of 7 pixels (1\farcs26). We use an optimized extraction technique that weights by the inverse-variance and the expected spatial profile of the source to maximize the signal-to-noise (S/N) of the extraction \citep{horne86}. As there was no continuum detected for any of our high-redshift sources, we used the wavelength-dependent spatial profile of the reference star in our mask to account for any seeing variations as a function of wavelength \citep[cf.,][]{song16,star17}.

\begin{figure}[ht] 
\centering
\includegraphics[width=\linewidth]{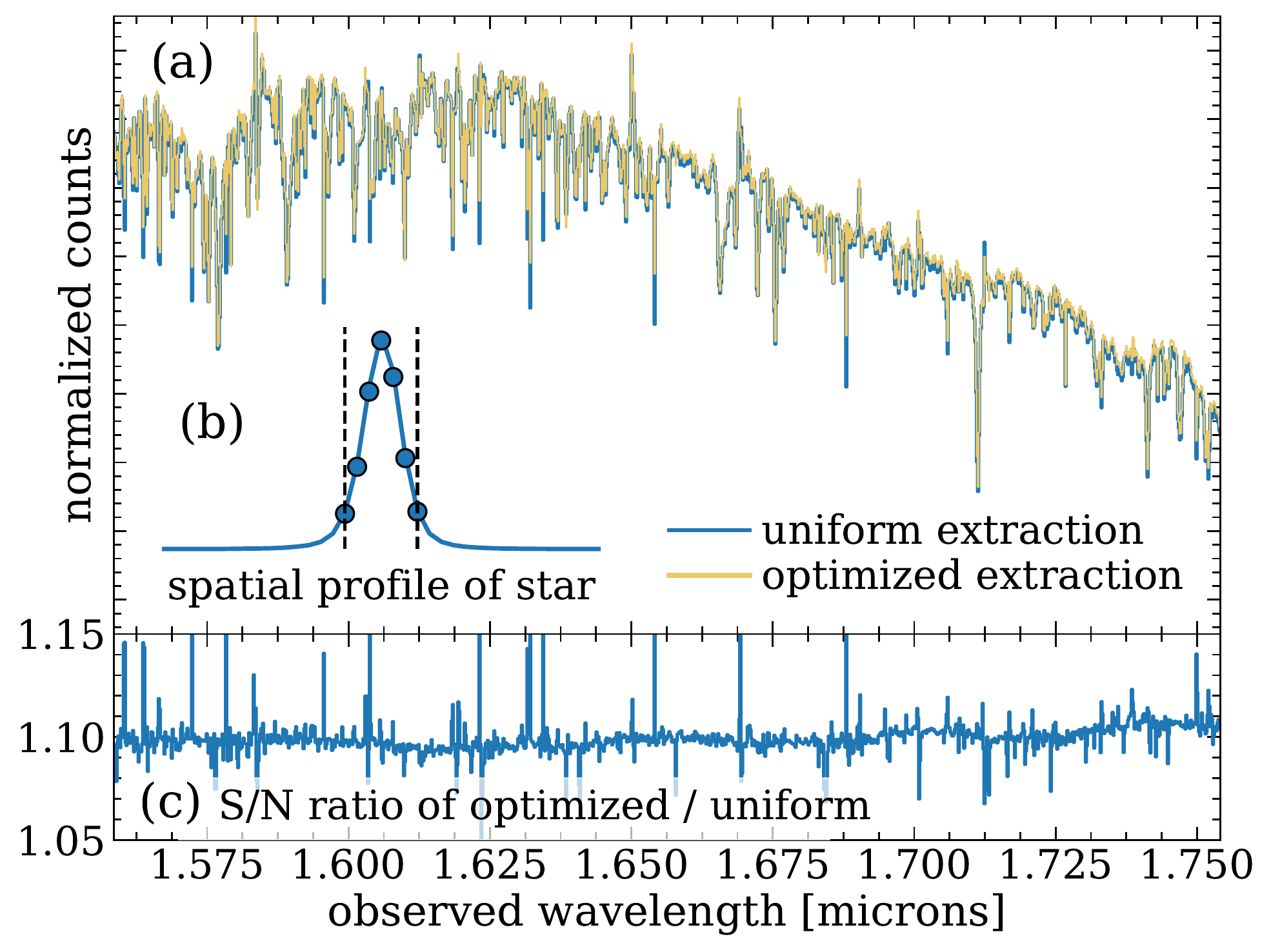}
\caption{Comparison of the optimized 1D spectral extraction technique versus traditional uniform extraction for the alignment star in the 2017 April 18 mask. The top panel shows (a) the spectrum extracted using a uniform aperture and the optimized method (as labeled in the plot inset) and (b) the weight used by the optimized process, defined within the 7-pixel extraction aperture, centered on the peak of emission (spatially). The bottom panel (c) shows the ratio of the S/N of the extracted spectra for the optimized method versus the uniform extraction.  There is an overall increase in recovered S/N for the optimized extraction. \label{fig:comp1d}}
\end{figure}

We applied the identical optimized extraction method for all sources and standard stars to provide consistency. Figure \ref{fig:comp1d} shows a comparison of the optimized method versus the traditional ``boxcar'' (uniform) extraction for the reference star in our 2017 mask. The spectra shown are pre-flux calibration and telluric correction. The inset, Figure \ref{fig:comp1d}b, shows the spatial profile of the reference star, along with the defined 7-pixel aperture and the series of weights, centered on the peak of emission, used to increase the S/N of the extracted spectra. The weights are made such that at each wavelength step they sum to unity.   Figure~\ref{fig:comp1d}c shows that the optimized-extraciton technique increases the S/N by only $\sim$10\%.   However, this is expected for spectra with  medium-to-high S/N, as the improvement in S/N is greatly enhanced for objects with much lower S/N \citep[see, e.g.,][]{horne86}.

To derive errors on the 1D spectra, we developed a method which takes advantage of multi-slit observations targeting faint objects (where some or many slits may result in non-detections). The method is described fully in Appendix~\ref{appsub:error}.  In brief, we chose regions spatially along the mask devoid of a signal or negative traces and extracted ``blank'' apertures following the same optimized extraction as used on our targets.  After gathering at least N $=$ 20 of these apertures, we determined the standard deviation of the distribution of the ``blank'' 1D spectra at each wavelength step as the error. The final 1D error spectrum was then used for all objects in the mask.  

For the absolute flux calibration of the spectra we followed two steps.  First, we corrected the data for telluric absorption and instrument response using the longslit (2014 data) or spectrophotometric `long2pos' (2017 data) observations of the standard star HIP 53735 (2014 data) and HIP 56147 (2017 data). Both stars have spectral type A0V, and were taken immediately before (2014 data) or in the middle of (2017 data) the observing block for each mask observation. As there were no $H$-band standard star frames taken on 2014 March 25, we used those taken on 2014 March 14 (but we refine the flux calibration in the second step, see below). 

Second, we used the reference star in each science mask to refine the absolute flux calibration.  This accounts for any variation in seeing or atmospheric transmission between the observation of the spectrophotometric standard and the science mask.  We first fit Kurucz models \citep{kuru93} to the fluxed spectrum of the reference star from our MOSFIRE mask, and use this to extrapolate the spectrum over the entire wavelength range covered by the {\it HST}/WFC3 F160W bandpass.  We then measured a synthetic F160W magnitude by integrating the MOSFIRE spectrum of the star with the F160W bandpass, and fit a scale factor to adjust this to match the (measured) total magnitude in the CANDELS photometry, using the catalog of Finkelstein et al.\ (in prep).   We then applied the same scale factor to the corrected MOSFIRE spectra for the galaxies in our sample to derive their absolute flux scaling.   We repeated this step for each mask.    We then measured the mean of the individual 1D spectra for each galaxy from each mask, weighting by the total exposure time of each mask. 

\section{Results} \label{sec:result}

\begin{figure*}[htb!]
\centering
\includegraphics[width=0.88\textwidth]{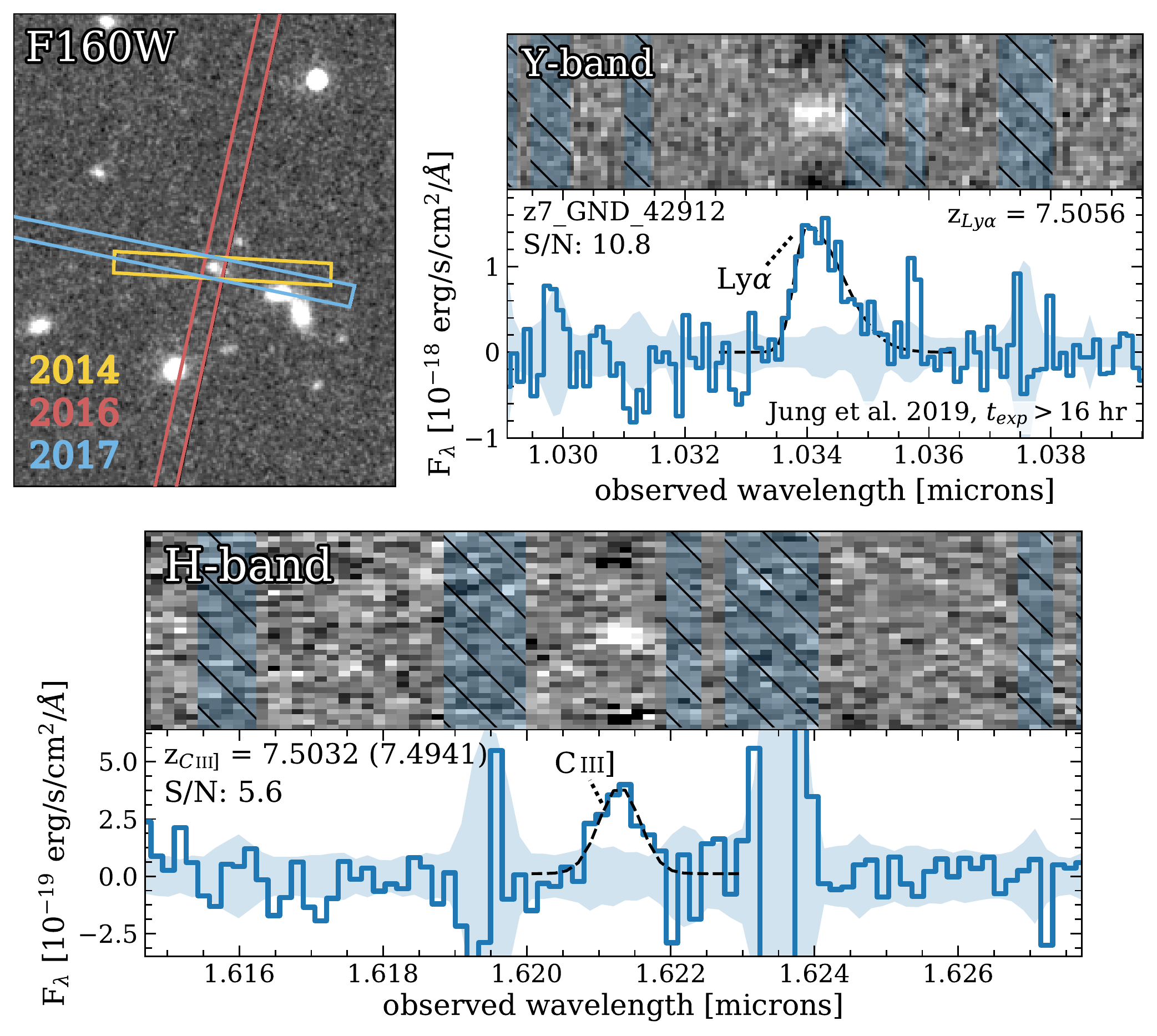}
\caption{{\it Top left:} A zoom in of the CANDELS/GOODS-N field in F160W band, centered on \gndone.  The different-colored slits indicate the different datasets and their respective positions on the target. {\it Top right:} $Y$-band MOSFIRE spectroscopy of \gndone\ from \citet{jung19}, showing the 2D S/N spectrum (where the S/N spectrum is the flux divided by the error; the blue hatched regions mask skylines), and the optically extracted 1D spectrum (solid blue line) with error (shaded blue region). \editone{The black dashed line shows a (skewed) Gaussian fit to the line.}  {\it Bottom:} $H$-band MOSFIRE spectroscopy showing the combined 2D smoothed S/N spectrum (top panel) and the optimally-extracted 1D spectrum from the combined dataset (bottom panel).  The blue shading is the same as in the top right panel. \editone{The black dashed line shows a Gaussian fit to the line .} \label{fig:hband}}
\end{figure*}

\gndone\ (n\'e z8\_GND\_5896, also known as FIGS\_GN1\_1292) is a bright (H$_{\text{160}}$ = 25.38, Finkelstein et al.\ in prep), highly star-forming galaxy at $z_{Ly\alpha}$ = 7.5078. The brightest detection in our $H$-band sample of high-redshift galaxies, \gndone\ was first spectroscopically confirmed via \lya\ emission by F13. 
Figure \ref{fig:hband} shows a portion of the  {\it HST}/WFC3 F160W image of the  CANDELS/GOODS-N field, including the MOSFIRE slit positions from the MOSFIRE 2013, 2014, and 2017 observing runs.   Figure~\ref{fig:hband} also shows the $Y$--band and $H$--band 2D and 1D combined spectroscopy of this galaxy.  

\subsection{Reanalysis of the Ly$\alpha$ Emission} \label{subsec:lya}

The top right of Figure \ref{fig:hband} shows the $>16$ hr $Y$-band 2D and 1D spectra from J19, centered on the observed \lya\ emission.  The top panel shows the 2D $Y$-band S/N spectrum (i.e., the image of the flux divided by the error), with skylines masked out. The bottom panel shows the corresponding  
1D $Y$-band flux spectrum with the blue-shaded regions denoting the error.
 
As discussed in F13, the profile of the \lya\ emission appeared symmetric, atypical (but not unseen) for \lya\ emission at higher redshifts \citep[e.g.,][]{dijk14,erb14}, where generally the blue half of the emission line has been absorbed and scattered by \hi\  in the IGM.  
The more recent analysis of J19 used $>$16 hr of  MOSFIRE $Y$-band data shows an asymmetric \lya\ line profile (although the uncertainties do not rule out that the line is symmetric). 
With this reported asymmetry, it could indicate a larger offset between the \lya\ emission and systemic redshift for this galaxy.  However,  this interpretation is not unique as some $z>6$ sources show both an asymmetric \lya\ line profile with a small offset from the systemic redshift \citep[e.g.,][]{star15a}.

Deep (40-orbit) {\it HST}/WFC3 G102 observations were taken for this galaxy as part of the Faint Infrared Grism Survey \citep[FIGS]{pirz17}.  Using a portion of the FIGS data, T16 showed that while the location of the \lya\ emission matched well between those observations and F13, the measured \lya\ line flux was a factor of $\sim$4 higher for the grism data (see Table 1 of T16). Using the full FIGS dataset, \citet{lars18} measured a \lya\ line flux consistent with those of T16. However, it is important to note that with the revised flux calibration for the 2013 April 18 $Y$-band data (J19), this discrepancy disappears and becomes consistent with the grism measurement, within errors.\footnote{\editone{In brief, the c.\ 2013 version of the MOSFIRE DRP provided different units (not in the original documentation) between the multi-object spectral frames and the longslit standard star frames. This has been corrected.}}
\editone{The remaining differences between the equivalent width of \lya, W$_{Ly\alpha}$, measurements for the two datasets arise from the different techniques applied to these independent data to calculate this value.}
Here, we adopt the \lya\ line flux and redshift from J19 (see Table~\ref{table:1.5}).

\subsection{Analysis of the \ion{C}{3}] Emission}

Based upon the \lya\ redshift, the expected location of the \ciii\ $\lambda\lambda$1907,1909 lines would be within a few hundred km s$^{-1}$ of 1.622 $\mu$m and 1.624 $\mu$m, respectively.  For each $H$-band dataset, we visually inspected the spectra in this wavelength range. The bottom panel of Figure \ref{fig:hband} shows 2D and 1D spectral region of the expected location of the \ciii\ lines.  We identified an emission line in this region at 1.6213 $\mu$m, which we label as one of the \ciii\ lines.  We fit a Gaussian to the 1D spectrum for the observed \ciii\ emission line, from which we derive a line flux of (2.63 $\pm$ 0.52) $\times10^{-18}$ erg s$^{-1}$ cm$^{-2}$, at S/N = 5.6.   We also inspected the spectrum over the full wavelength range, but failed to identify any other candidate emission line. 

\begin{figure}[htp]
\centering
\includegraphics[width=\linewidth]{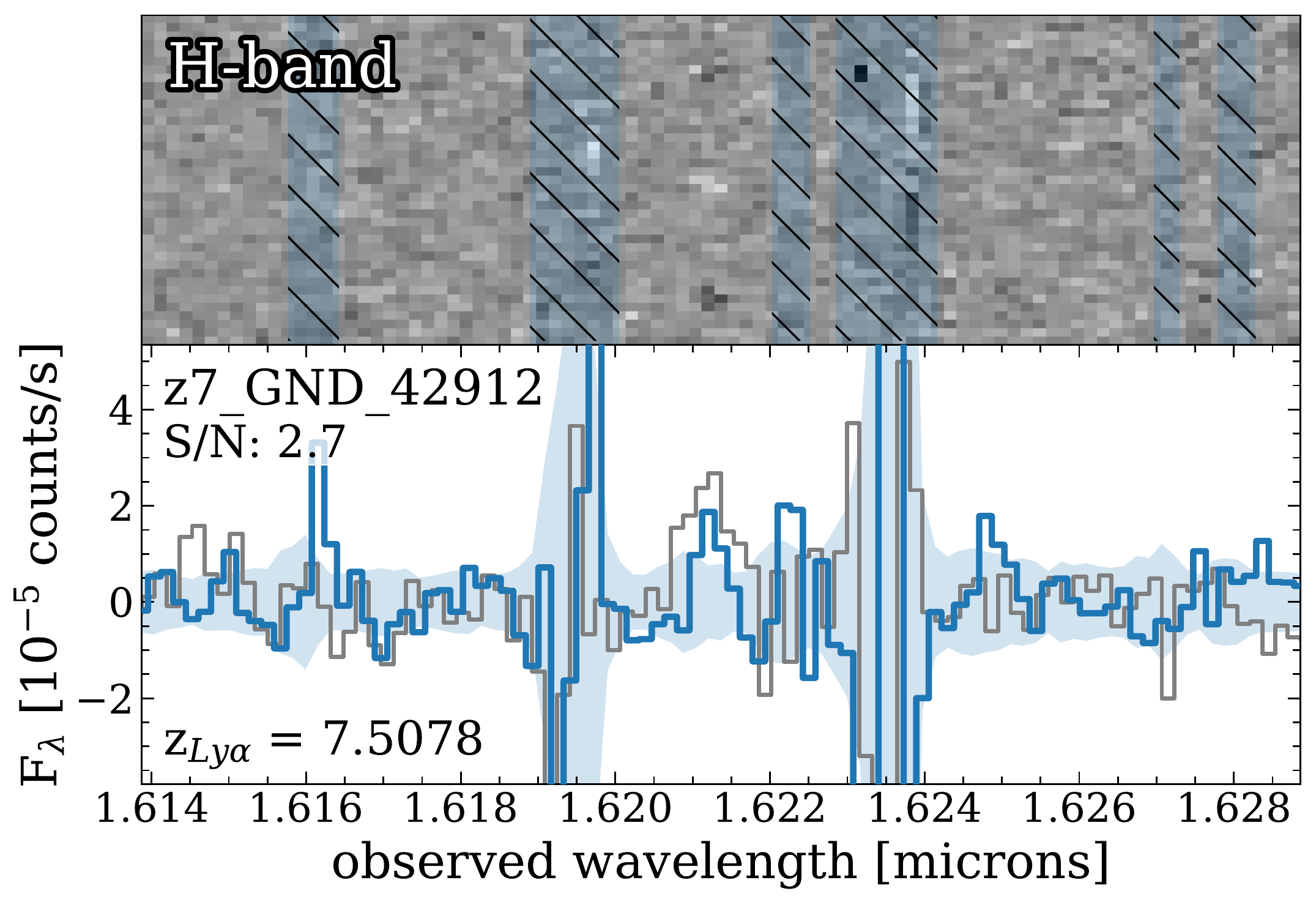}
\caption{$H$-band MOSFIRE spectroscopy from the 2016 May 4 data of \gndone\ showing an independent detection of \ciii\ emission. The data are not flux calibrated. The top panel shows the 2D smoothed S/N spectrum, with the blue hatched regions masking out skylines; the bottom panel shows the optimally-extracted 1D flux spectrum, shaded blue regions indicating the error. Overplotted in grey is the 1D co-added and flux-calibrated spectrum from Figure \ref{fig:hband}, scaled to be visible. \label{fig:zitrin-gndone}}
\end{figure}

In addition, we verified the detection of this line using data taken for this galaxy from an independent MOSFIRE program from 2016 May 4 (PI Zitrin, see Table \ref{table:1}).   We reduced the data following the same steps as discussed above in~\sref{subsec:flux-cal}. However, these data lacked a telluric standard observation.  Therefore, we make no attempt to calibrate the spectrum, but these data provide a robust, independent detection of the line.  Figure \ref{fig:zitrin-gndone} shows the $H$-band 2D smoothed S/N spectrum and 1D optimally-extracted flux spectrum of the 2016 data..  We identify an emission line at the same wavelength as the detected \ciii\ with a S/N of 2.7.   This provides additional confidence in the detection of this emission line. 

In order to determine the systemic redshift, we need to assign an identification to this line.  As mentioned before, \ciii\ is represented as a doublet\footnote{Strictly speaking, \ciii\ is a line pair \citep[see, e.g.,][]{star14}, a combination of one forbidden and one semiforbidden transition -- namely [\ciii\ $\lambda$\editone{1906.68} and \ciii\ $\lambda$1908.73.  When applicable, we will reference the individual lines using the notation [\ciii$\lambda$1907 and \ciii$\lambda$1909, respectively.} with rest-frame separation of $\sim$2\ang; which for the redshift of this galaxy corresponds to an observed separation of $\sim$16\ang. Due to the wavelength of the observed line, in both cases ([\ciii\ $\lambda$1907 or \ciii\ $\lambda$1909) the expected location of the other line falls on one of the two strong sky lines adjacent to the emission feature (see the $H$-band 2D spectrum in Figure \ref{fig:hband}). This prevents us from making a robust identification, as the other line is unobservable in both cases.   We discuss this complication and its implications in more detail in~\sref{subsec:sysz}.

In addition to \ciii, the \siiii\ $\lambda$1883,1892 doublet would also fall in the MOSFIRE $H$--band spectrum.  This line is observed in some high redshift galaxies, with a typical flux ratio of (\siiii\ $\lambda$1883)/([\ciii+\ciii) = 0.1--0.3 \citep[e.g.,][]{star14,stei16,patr16,berg18}.    We therefore inspected the 2D spectra for signs of the these lines, but failed to identify any line at the expected location of \siiii.  We forced a Gaussian fit for \siiii\ lines at the expected wavelengths, using the location and width of the \ciii\ line. This places a formal 2$\sigma$ upper limit of \siiii\ $\lambda$1883 $<$ 0.92 (1.08)\,$\times10^{-18}$ erg s$^{-1}$ cm$^{-2}$, assuming the observed emission is [\ciii\ $\lambda$1907 (\ciii\ $\lambda$1909).  These results imply a (\siiii\ $\lambda$1883)/(\ciii\ $\lambda$1907) ratio of $<$0.35 (2$\sigma$), and a (\siiii\ $\lambda$1883)/([\ciii+\ciii) ratio of $<$0.21 (2$\sigma$).  We discuss this in more detail below in \sref{subsec:models}.

\begin{figure*}[hbt]
\centering
\includegraphics[width=0.8\linewidth]{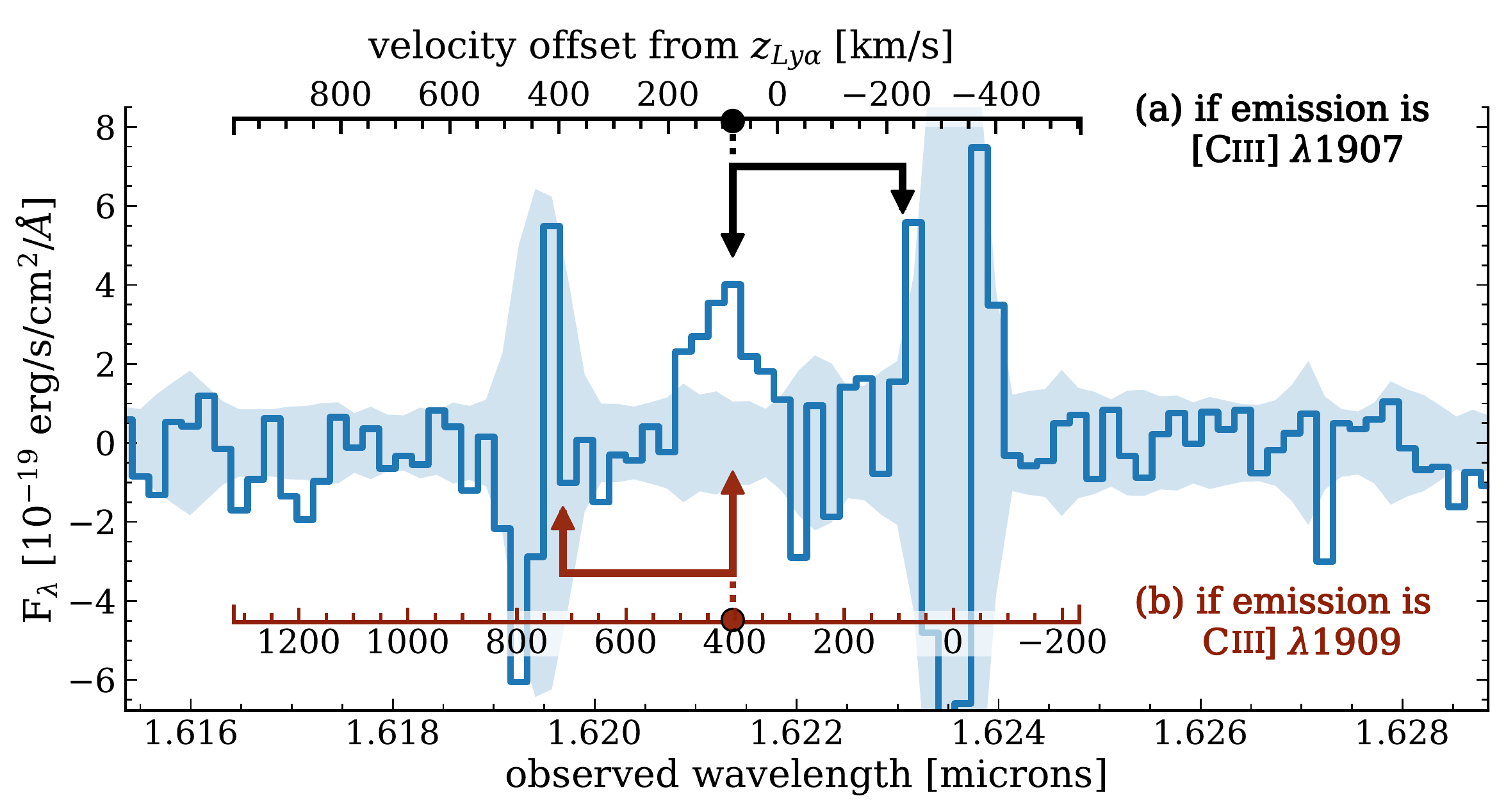}
\caption{Systemic redshift determination for \gndone\ for the case that the detected emission line is (a) [\ciii\ $\lambda$1907 or (b) \ciii\ $\lambda$1909 as a function of both wavelength and velocity offset from $z_{Ly\alpha}$. The black and red arrows indicate the expected location of both lines of the \ciii\ emission doublet for case a:  the identified line is [\ciii\ $\lambda1907$, and case b: the identified line is \ciii\ $\lambda1909$, respectively.  The inset axis shows the velocity offset of \lya\ from systemic for each possible \ciii\ identification, and the arrows show the expected location of the other line of the doublet (in both cases they would fall on sky emission lines).  \label{fig:which-sysz}}
\end{figure*}

\subsection{Implications for Nebular Emission from the \\IRAC [3.6]--[4.5] Color}

Apart from \ciii, this galaxy also shows evidence of additional, strong emission lines in the colors of the galaxy's SED.  From F13, the reported IRAC color was found to be very red, with [3.6]--[4.5] = 0.98 mag.     Using updated photometry, this color is revised downward to [3.6]--[4.5] = $0.77^{+0.23}_{-0.28}$ mag (Finkelstein et al., in prep; although consistent with the photometric errors of F13).  \gndone\ has a high SFR of \editone{$180^{+20}_{-50}$ $M_\odot$~yr$^{-1}$} from
the analysis of the full SED (J19).   The interpretation of the red IRAC color suggests the galaxy has strong nebular emission from \hb\ + \oiii\ in the 4.5 $\mu$m band with an inferred restframe \oiii\ $\lambda$5007 equivalent width (W$_{\text{[OIII]}}$) of $\sim$600\ang~(F13, Finkelstein et al.\ in prep; following the prescription of \citealt{papo01}).  F13 used the strength of this emission to constrain the metal abundance of \gndone\ ($\simeq$0.2--0.4~$Z_\odot$) and its relatively young inferred age (10 Myr). In the discussion below, we will use both the implied \oiii\ equivalent width from the red IRAC color and our \ciii\ measurement to study the ionization conditions and metallicity of this galaxy (see~\sref{subsec:models}).

\section{Discussion} \label{sec:discuss}

\subsection{Systemic Redshifts and Implications for \lya} \label{subsec:sysz} 

Unlike \lya\ (which is subject to strong resonant scattering), the \ciii\ nebular emission line provides a measure of the systemic redshift of a galaxy.  This can be used to measure the velocity offset of \lya\ (\vlya) from the systemic, which then provides insight into the kinematics of a galaxy \citep[e.g.,][]{erb14,star17} and information about the \lya\ escape fraction \citep[e.g.,][]{hayes15}.

In the case of \gndone, we detect one of the lines in the \ciii\ doublet with a S/N of 5.6.  In order to determine which line we have detected, we took into account the implied \vlya\ derived from the systemic assuming each of the possible lines. Figure \ref{fig:which-sysz} shows the expected location of the other line assuming the detected emission is (a) [\ciii\ $\lambda$1907 or (b) \ciii\ $\lambda$1909. As discussed above (\sref{sec:result}), in both cases the other line of the \ciii\ doublet would be located in a region with strong night sky emission.  The velocity offset for the case that the line is \ciii\ $\lambda$1909 is \vlya\ = \editone{$410\pm27$ km s$^{-1}$}.   The velocity offset for the case that the line is [\ciii\ $\lambda$1907 is \vlya\ = \editone{$88\pm27$ km s$^{-1}$}.\footnote{\editone{Due to the asymmetric fit of the \lya\ emission, there exists an additional possible systematic uncertainty associated with the centering of this emission line, where a more symmetric fit would result in a $\sim$70\% increase in the measured \vlya.}}
These reported velocity offsets have been corrected to the heliocentric frame, \editone{with the \lya\ and \ciii\ datasets having mean corrections of -7.79 km/s and -11.57 km/s, respectively \citep{mcli11,song14,chon13}. (It is important to note that while we apply this correction to the velocity offsets, we do not correct the redshifts listed in Table \ref{table:1.5}.})

\begin{figure}[thp]
\centering
\includegraphics[width=\linewidth]{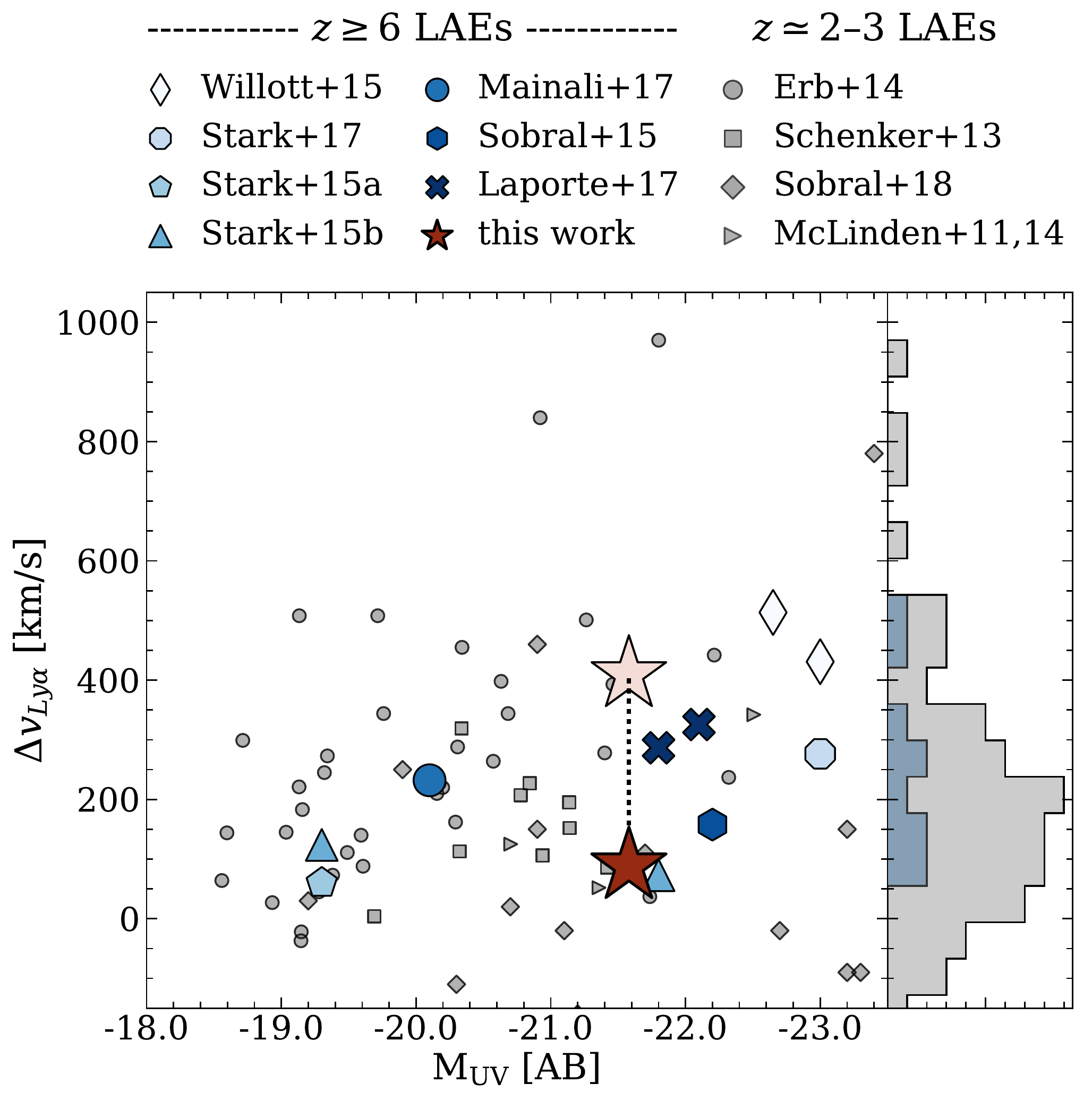}
\caption{A compilation of velocity offsets for galaxies at $z\geq6$ with both \lya\ and a nebular line (which provides the systemic redshift of the source). In addition, we have included a sample of galaxies  at $z\approx2-3$ from \citet{erb14}, \citet{schen13}, \citet{sobr18}, \citet{mcli11}, and \citet{mcli14}, respectively.  The stars represent \gndone\ where the dark (light) star assumes the detected emission is [\ciii\ $\lambda$1907 (\ciii\ $\lambda$1909).
\label{fig:delv-muv-z}}
\end{figure}

The value of \vlya\ has implications for kinematics, outflows, and geometry in \gndone.   From a study of mid-redshift (z $\simeq$ 2--3) star-forming galaxies, \citet{erb14} suggest a correlation exists between \vlya\ and UV luminosity, while an anti-correlation exists between \vlya\ and \lya\ equivalent width.   \citeauthor{erb14}\ argue these are a consequence of changing the column density, covering fraction, and/or velocity dispersion of gas within the ISM -- all effects found to impact the emergent profile of \lya\ emission \citep{verh06,dijk14}.    The distribution of \vlya\ is also expected to decrease for increasing redshift \citep{schen13,chou15,star15a}, implying an emergence of harder ionizing radiation fields that can blow out a substantial fraction of neutral gas in the ISM. This clearing of gas would mean that \lya\ photons can escape into the (neutral) IGM (this has additional implications for the leakage of ionizing Lyman continuum photons, e.g. \citealt{kimm19}).  

Figure~\ref{fig:delv-muv-z} shows a compilation of rest-frame UV absolute magnitude versus \vlya\ for galaxies in the literature at $z > 6$ and $z\sim 2-3$.  One point immediately evident from the figure is that few galaxies have high \vlya, with a mode of the distribution (i.e., the typical value) of $\simeq$200 km s$^{-1}$.     At $z\sim 2-3$, studies of LAEs find very few galaxies at all UV luminosities with \vlya\ $>$300~km s$^{-1}$.    At $z > 6$, the only galaxies with high \vlya~$>$350~km~s$^{-1}$ have $M_\text{UV} < -22.5$~mag  \citep[e.g.,][]{will15,star17}.   All galaxies with luminosities fainter than $M_\text{UV} > -22.4$~mag, have lower \vlya\ \citep[see,][]{star17}. 

\begin{deluxetable*}{cccccccc}[thp]
\tablecaption{\label{table:1.5}Measurements for \gndone}
\tablecolumns{8}

\tablehead{\colhead{$z_{Ly\alpha}$} & \colhead{$z_{sys}$} & 
\colhead{Line} & \colhead{$\lambda_0$} & \colhead{$\lambda_{obs}$} &
\colhead{$f_{line}$} & \colhead{W$_{line,0}$} & \colhead{Reference} \\
\colhead{} & \colhead{} & \colhead{} & \colhead{{\small (\ang)}} & 
\colhead{{\small ($\mu$m)}} & \colhead{{\small (10$^{-18}$ erg$^{-1}$ s$^{-1}$ cm$^{-2}$)}} & 
\colhead{{\small (\ang)}} & \colhead{} \\
\colhead{(1)} & \colhead{(2)} & \colhead{(3)} & \colhead{(4)} & \colhead{(5)} & \colhead{(6)} & \colhead{(7)} & \colhead{(8)}}

\startdata
7.5078 $\pm$ 0.0004 & \dotfill~ & \lya & 1215.67 & \editone{1.03427 $\pm$ 0.00005} & 2.64 $\pm$ 0.74$\,$\tablenotemark{a} & 7.5 $\pm$ 1.5$\,$\tablenotemark{b} & \citealt{fink13} \\
7.512  $\pm$ 0.004 & \dotfill~ & \lya & 1215.67 & 1.035 $\pm$ 0.005 & 10.6 $\pm$ 1.9 & 49.3 $\pm$ 8.9\,\tablenotemark{b} & \citealt{tilvi16} \\
7.510 $\pm$ 0.003 & \dotfill~ & \lya & 1215.67 & 1.033 $\pm$ 0.004 & 11.0 $\pm$ 1.7 & & \citealt{lars18} \\
\editone{7.5056 $\pm$ 0.0007}$\,$\editone{\tablenotemark{c,f}} & \dotfill~ & \lya & 1215.67 & \editone{1.0340 $\pm$ 0.0001} & 14.6\,\tablenotemark{f} $\pm$ 1.4 & 33.2 $\pm$ 3.2\, \tablenotemark{b}  & \citealt{jung19}\\
     & \editone{7.5032 $\pm$ 0.0003} & [\ciii & \editone{1906.68} & \editone{1.62129 $\pm$ 0.00006} & 2.63 $\pm$ 0.52 & 
     \multirow{2}{*}{16.23 $\pm$ 2.32\,\tablenotemark{d}} & This work \\
     & \dotfill~ & \ciii & 1908.73 & ~\dotfill~ & ~~1.74 $\pm$ 0.35$\,$\tablenotemark{e} & & This work \\
     & \dotfill~ & \siiii & 1882.47 & ~\dotfill~ & $<0.924~ (2\sigma)$ & & This work \\
     & \dotfill~ & \siiii & 1892.03 & ~\dotfill~ & ~~~~~~~\dotfill~~~~~~~  & & This work \\
\enddata

\tablecomments{(1) Redshift derived from \lya. (2) Systemic redshift derived from [\ciii. (3) Spectroscopic line measured. (4) Restframe wavelength of line. (5) Observed wavelength for line. (6) Line flux. (7) Restframe equivalent width. (8) Reference for a given row of measurements.}
\tablenotetext{$a$}{\editone{The previous lower line flux resulted from an earlier discrepancy, which has been resolved and updated as described in \citealt{jung19} (see also Section~\ref{subsec:lya}).}}   
\tablenotetext{$b$}{\editone{The differences in \lya\ equivalent width in the literature result from the revised \lya\ line flux (see \citealt{jung19} and Section~\ref{subsec:lya}) and updates to the CANDELS photometry (Finkelstein et al., in prep).  There are also additional differences that stem from the methods used to analyze {\it HST}/grism data (\citeauthor{tilvi16}) compared to the MOSFIRE data (\citeauthor{jung19}).}}
\tablenotetext{$c$}{\editone{Redshift derived from asymmetric fit to \lya\ profile.}}
\tablenotetext{$d$}{\editone{Based on the identification of the emission line as [\ciii\ $\lambda$1907.}}
\tablenotetext{$e$}{\editone{Line flux inferred by using a [\ciii\ / \ciii\ ratio of 1.5 $\pm$ 0.1. Note ratio also has a systematic uncertainty of 7\% depending on electron density and temperature, see \S~\ref{subsec:sysz}.}}
\tablenotetext{$f$}{\editone{Values for \lya\ used in the analysis in this work.}}
\end{deluxetable*}

\gndone\ has M$_\text{UV}=-21.58$ (J19). Comparing \gndone\ to other galaxies in Figure~\ref{fig:delv-muv-z}, we favor the interpretation that the line is [\ciii\ $\lambda$1907.  This is consistent with the distribution of other galaxies, and implies the lower value of \vlya\ for \gndone\ (\editone{$\vlya = 88\pm27$ km s$^{-1}$}), corresponding to a systemic redshift \editone{$z_{sys} = 7.5032 \pm 0.0003$}. In this case, as described above, a harder ionization field would likely be responsible for clearing \hi\ around the galaxy, allowing \lya\ to escape closer to $z_{sys}$ \citep{schen13,star15a,star17}.  

Additional evidence in support of the [\ciii\ $\lambda$1907 interpretation comes from studies of ``green pea'' galaxies at $z\sim 0.1-0.3$. 
Using a sample of 43 green pea galaxies,\citet{yang17} find a correlation between 
\lya\ equivalent width and \lya\ escape fraction.   Furthermore, they find that galaxies with lower \vlya\ and lower dust attenuation show larger values of \lya\ escape fraction.   Given the relatively high \lya\ equivalent width and lower dust attenuation of \gndone\ (F13, J19),  this provides additional support for a lower \vlya, and supports  that  observed emission line is [\ciii\ $\lambda$1907.

\begin{figure}[tbp]
\centering
\includegraphics[width=\linewidth]{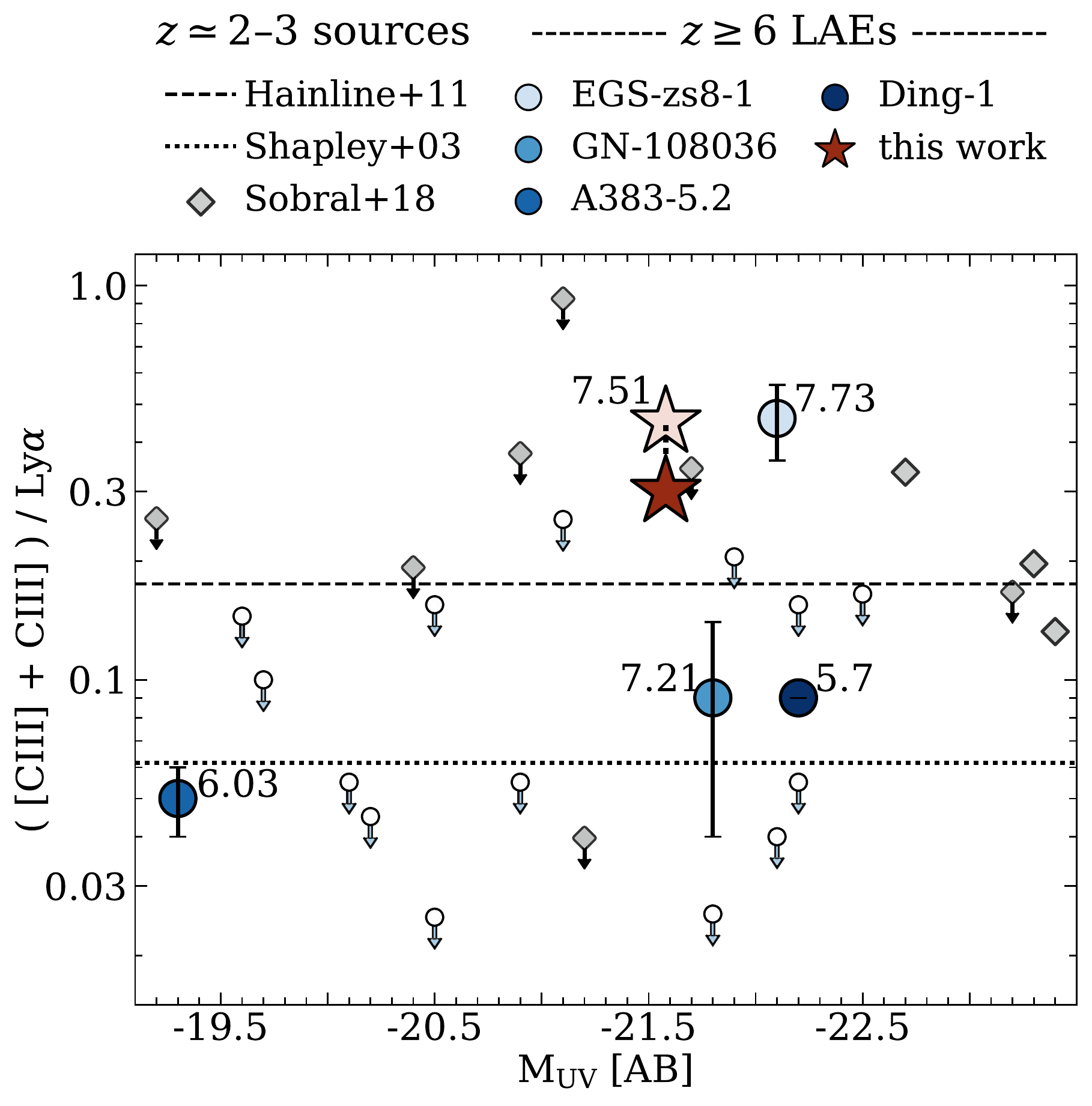}
\caption{All measured $z\geq6$ galaxies with spectroscopic measurements in both \lya\ and \ciii\ from the compilation in \citet[and references therein]{matt17}. 1$\sigma$ upper limits are shown as open circles with arrows, while the detections are colored circles. The $z_{Ly\alpha}$ of the detections are shown next to the points. The stars represent \gndone\ where the dark (light) star assumes the detected emission is [\ciii\ $\lambda$1907 (\ciii\ $\lambda$1909).  In addition, we have included a sample of $z\approx2-3$ LAEs from \citet{sobr18}, represented by the grey diamonds and 1$\sigma$ diamond upper limits. The dashed line represents the $z\sim3$ UV-selected AGN composite from \citet{hain11} and the dotted line represents the $z\sim3$ LBG composite from \citet{shap03}. \label{fig:ciii-lya}}
\end{figure}
 
However, while we identify the emission line as [\ciii\ $\lambda$1907, we are unable to rule out that the detected line is \ciii\ $\lambda$1909.  We therefore consider both possibilities in the discussion that follows.      Future observations of other nebular emission lines will test this interpretation (e.g., from {\it JWST}, see \sref{subsec:jwst} below). 

Lastly, we estimated the total flux of the combined [\ciii\ $\lambda$1907 + \ciii\ $\lambda$1909 lines in order to compare to previous results and models.  Many previous studies are unable to resolve these lines.  The \ciii\ doublet is also blended in the spectra output from some radiative transfer codes (e.g., {\sc Cloudy}).   To estimate the total flux,  we use models for the [\ciii/\ciii\ ratio (see~\sref{subsec:models}), combined with the line flux we measure.   This is similar in practice to methods used in the study of other galaxies at $z >$ 1 \citep[e.g.,][]{sand16,mase17} in the case that only part of the \ciii\ doublet is identified \citep[e.g.,][]{star15a}.  For a range of gas densities 10$^2$--10$^4$ cm$^{-3}$, the [\ciii/\ciii\ ratio varies from $\sim$1.5--1.2 \citep{mase17}.  
 For an electron density of 300 cm$^{-3}$ (see~\sref{subsec:models}), our models return a ratio of 1.5 \editone{(with a range of 0.1, depending upon electron density and temperature}) for all metallicities considered ($Z$ = 0.05--0.5 $Z_\odot$).
This is consistent with previous work to within 10\% \citep[e.g.,][]{star15a,mase17}.  Therefore, we adopt this ratio,  ([\ciii~$\lambda$1907)/(\ciii~$\lambda$1909) = 1.5 for all calculations.    The detected emission line has a flux of $F_\mathrm{Line}$ = $(2.6 \pm 0.5) \times 10^{-18}$~erg s$^{-1}$ cm$^{-2}$ (independent of line identification).   For the case that the detected emission line is [\ciii\ $\lambda$1907, this yields a total line flux of $(4.4 \pm 0.8) \times 10^{-18}$~erg s$^{-1}$ cm$^{-2}$, corresponding to a rest-frame \wciii = $16.23\pm2.32$. 
These measurements are summarized in Table \ref{table:1.5}. For the alternative case that the detected emission line is \ciii~$\lambda$1909, the total line flux would be higher by approximately 50\%,   $(6.6 \pm 1.3) \times 10^{-18}$~erg s$^{-1}$ cm$^{-2}$, and would correspond to \wciii = $25.09\pm2.32$.

Using the combined total line flux for  [\ciii+\ciii, Figure \ref{fig:ciii-lya} compares the relationship between the (total) \ciii\  and \lya\ emission for \gndone\ to other galaxies with line measurements at $z \geq$ 6 \citep[using][and references therein]{matt17}.  For \gndone, we measured a ([\ciii+\ciii)/\lya\ ratio of 0.30 $\pm$ 0.04.  Most measurements in the literature provide only  upper limits on \ciii\ (shown in the figure as arrows) while there are a handful of detections (shown in the figure as circles).    \gndone\ has one of the highest \ciii/\lya\ ratios yet measured during this epoch (and this would be even higher in the case where the line is \ciii\ $\lambda$1909).  (We note that using the {\it HST} grism \lya\ line flux (see Table~\ref{table:1.5}) would \textit{increase} this ratio by $\simeq$40\%.) From these results, there does not seem to be any significant trend between the ratio of ([\ciii+\ciii)/\lya\ and redshift or M$_{\text{UV}}$, although the sample size is still too small for robust conclusions.  \editone{This apparent randomness is not unexpected, as the emergent \lya\ emission measured in these sources may be subject to different effects.  Therefore, it is not immediately clear if the ([\ciii+\ciii)/\lya\ ratio is indicative of the ionization state of these galaxies, however this ratio could be investigated as an interesting probe of these sources.} 

\subsection{Physical Interpretation of the Emission Lines} \label{subsec:models}

For \gndone, the ([\ciii+\ciii)/\lya\ flux ratio and red [3.6]--[4.5] color suggest high ionization.  Several studies have argued that these red IRAC [3.6]--[4.5] colors for galaxies in the reionization era imply strong nebular emission from \hb\ +\oiii, redshifted into the IRAC bandpasses \citep[e.g.,][]{smit15,robo16,star17,matt17}.  This is the case for \gndone, with [3.6]--[4.5] = 0.77~mag, which implies a rest-frame \oiii\ equivalent width (W$_{\text{[OIII]}}$) of $\simeq\,$600~\AA\ (see above).  As has been shown at lower redshifts ($z\sim2$), higher values of W$_{\text{[OIII]}}$ have been found to correlate closely with an increasing ionizing efficiency \citep{tang18}. This may be responsible for ionizing a significant fraction of the ISM, therein ionizing a substantial amount of \hi\ and allowing a higher \lya\ escape fraction. 

We investigated the properties of \gndone\ by calculating the nebular emission spectrum expected assuming different stellar population (SP) models, with variable ionization, Hydrogen gas density, stellar and nebular metallicities, and other parameters \citep[motivated by analyses of other galaxies in the literature, e.g.,][]{inoue11,stan14,sand16,stei16,jask16,star17,strom17,byler17,sobr18}. 
%
%
To study model-dependent systematic differences in the nebular emission spectrum, we used a variety of SPs that also vary the IMFs (including variations to the upper mass cut-off of the IMF, M$_{\text{IMF}}^{up}$) in order to test a large range of available parameter-space \citep[e.g.,][]{stei16}.

\begin{deluxetable*}{clccc}[!tp]
\tablecaption{\label{table:2}Full List of \cloudy Parameters Used in Simulations.}
\tablecolumns{5}
\tablehead{
\multicolumn{2}{c}{Parameter} & \colhead{\bpass SPs} & \colhead{Starburst99 SPs} & \colhead{AGNs}}
\startdata
(1) & $Z_{*}$ [$Z_{\odot}$] &  0.1, 0.2, 0.3, 0.5 & 0.05, 0.1, 0.4 & \dotfill \\
 (2) & $Z_{\rm neb}$ [$Z_{\odot}$] & $= Z_{*}$ 0.3, 0.5 & $= Z_{*}$ 0.3, 0.5 & 0.05, 0.1, 0.2, 0.3, 0.4, 0.5 \\
 (3) & $n_H$ [cm$^{-3}$] & 300 & 300 & 300 \\
(4) & log\,U & [-3.5, -1.5], steps of 0.2 & [-3.5, -1.5], steps of 0.2 & [-3.5, -1.5], steps of 0.2 \\
(5) & $\Omega/4\pi$ & 1.0 ($\Omega=4\pi$) & 1.0 ($\Omega=4\pi$) & open ($\Omega\ll4\pi$) \\
(6) & M$_{\text{IMF}}^{up}$ [M$_{\odot}$] & 100, 300 & 100 & \dotfill \\
\multirow{4}{*}{(7)} & \multirow{4}{*}{$\alpha$} & & $-1.30$; ~[0.1, 0.5) M$_{\odot}$ & \multirow{4}{*}{\texttt{table agn}\tablenotemark{a}} \\
	& & $-1.30$; ~[0.1, 0.5) M$_{\odot}$ & 
    	\multirow{3}{*}{\begin{math}
        \begin{rcases*}
          	-2.30; \\
        	-2.00; \\
            -1.70; \\
        \end{rcases*} 
        \end{math}[0.5, M$_{\text{IMF}}^{up}$] M$_{\odot}$} & \\
    & & $-2.35$; ~[0.5, M$_{\text{IMF}}^{up}$] M$_{\odot}$ & & \\
    & & & & \\
(8) & SFH [M$_{\odot}$ yr$^{-1}$] & continuous; 1.0 & continuous; 1.0 & \dotfill \\
(9) & Age [Myr] & 10, 30, 100 & 10, 30, 100  & \dotfill \\
\enddata
\tablecomments{(1) Stellar metallicity of the SPs, $Z_{*}$. (2) Nebular metallicity used for each $Z_{*}$ and AGN model. (3) Total Hydrogen density. (4) Ionization parameter ranging from $-3.5$ to $-1.5$ in increments of 0.2. (5) Covering factor, where 1.0 defines a closed geometry. (6) IMF upper mass limit for the SPs. (7) Powerlaw slopes of the IMFs and the AGN models. (8) Star formation histories. (9) Age of the SPs.}
\tablenotetext{$a$}{The continuum shape given by the \texttt{table\,\,agn} command from \cloudy is described in Table 6.3 of Hazy, the \cloudy documentation.}
\end{deluxetable*}

\begin{figure*}[ht]
\centering
\includegraphics[width=0.86\textwidth]{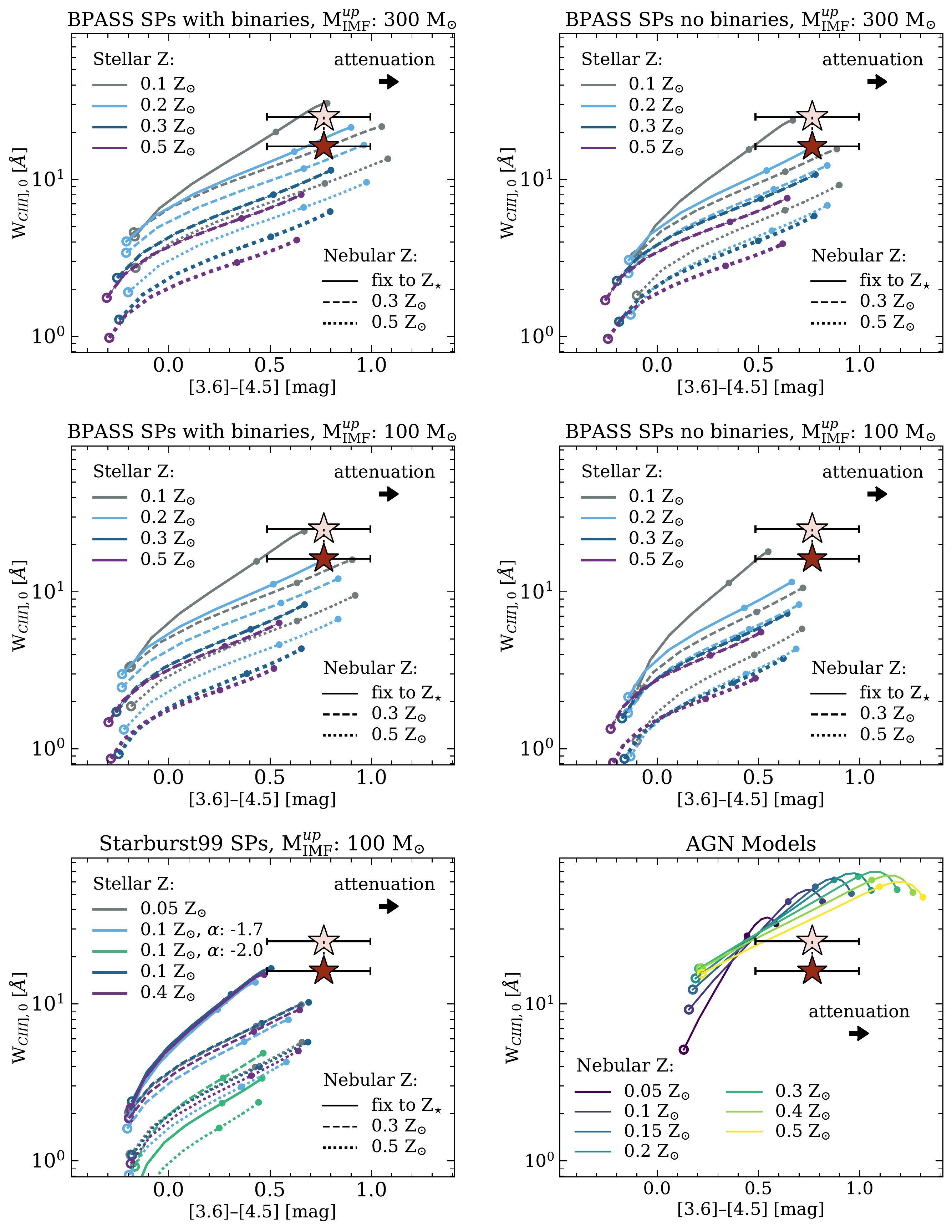}
\caption{ Restframe equivalent with of the \ciii\ doublet versus IRAC color.  The large stars represent \gndone\ where the dark red (light red) star assumes the detected emission is [\ciii\ $\lambda$1907 (\ciii\ $\lambda$1909).   The lines show models as a function of ionization parameter, stellar metallicity, and nebular metallicity for the single and binary \bpass SPs with M$_{\text{IMF}}^{up} =$ 100, 300 M$_{\odot}$ and the S99 SPs with M$_{\text{IMF}}^{up} =$ 100 M$_{\odot}$.   In the last panel, the \cloudy AGN models are shown, following the parameters listed in Table \ref{table:2}.    Each line shows the full range of ionization parameter explored (-3.5 to -1.5), with -3.5 indicated by the open circles and -2.5, -1.5 by the small circles. \editone{All of the SP models have continuous star formation with ages of 10 Myr}. The IRAC color was measured by redshifting the model spectra to $z_{sys}$ assuming the detected emission is [\ciii\ $\lambda$1907 (but using the redshift for \ciii\ $\lambda$1909 does not change the results). The solid black arrow shows the effect of dust attenuation using the reported $E(B-V)$ for \gndone\ as described by \citet{calz00}. \label{fig:cloudy-ciii-irac}}
\end{figure*}

We used the Binary Population and Spectral Synthesis v2.0 \citep[{\sc bpass};][]{eldr17} and Starburst99 \citep[S99;][]{leit14} models with continuous star formation histories (SFHs) of 1 M$_{\odot}$ yr$^{-1}$ with similar metallicity to \gndone\ (0.2--0.4 $Z_{\odot}$; based upon F13 values), varying the age from 10, 30, and 100 Myr (an age of 10 Myrs is within the age range derived from fitting models to the broad-band colors for this galaxy by \citealt{fink15} and J19). In order to test different models for the photospheres of massive stars and to allow the effects of binary stellar populations, we used both the single-star and binary-star SP models provided by {\sc bpass}.   The S99 SPs use Geneva 2012/13 stellar tracks assuming zero rotation, and cover a slightly different range of metallicity compared to the \bpass models (see Table~\ref{table:2}).  

We used the radiative transfer and photoionization microphysical code \cloudy v17.0 \citep{ferl17} to produce the nebular emission spectrum for each SP.  We modeled both the case where the nebular gas metallicity is fixed to be the same as the SP, and also the case where the nebular gas metallicity is allowed to vary from 0.3--0.5~$Z_\odot$ independent of the metallicity of the SP.  To normalize the ionizing continuum, we chose a range of ionization parameters, defined as the ratio of the number density of ionizing photons to the number density of the gas,  (U $\equiv n_\gamma$/$n_H$), running from log\,U = $-$3.5 to $-$1.5 in steps of 0.2~dex, assuming a covering factor ($\Omega/4\pi$) of 1. Per the methods followed in \citet{stei16}, we set the total gas density ($n_H$) to 300 cm$^{-3}$.   We assumed no attenuation by dust in the \cloudy models however we postprocess the model outputs to include dust attenuation assuming a foreground dust screen \citep{calz00} with the reported color excess $E(B-V)=0.22$ for \gndone\ (J19). \editone{We assume Solar elemental abundances in all of our modeling, but see below.}  Table~\ref{table:2} lists the full range of models and parameters used in our \cloudy modeling. 

In order to investigate the range of model parameter space that can reproduce the observed properties of \gndone, we focus on the total \ciii\ restframe equivalent width (W$_{\text{CIII]}}$; see Table \ref{table:1.5}) and the observed IRAC [3.6]--[4.5] color, which provides a measure of the \hb\ + \oiii\ equivalent width.  The values for the implied W$_{\text{[OIII]}}$ $\sim 600$\ang\ and total W$_{\text{CIII]}}$ are consistent with the relations seen in low metallicity dwarf galaxies at low redshifts, which show correlations between these quantities, and favor high ionization and lower metallicity \citep{senc17}.  We therefore expect these conditions may also exist in \gndone.  For each \cloudy output, we redshifted the spectrum to \editone{$z_\mathrm{sys} = 7.5032$}, and integrated them with the IRAC 3.6 and 4.5~\micron\ bandpasses \citep[following][]{papo01}.

Figure \ref{fig:cloudy-ciii-irac} shows the various \cloudy model results as a function of ionization parameter (for fixed model age of 10 Myrs) compared to the measurements for \gndone, assuming both \ciii\ line identifications, as in Figure~\ref{fig:delv-muv-z}.  Models with older stellar-population ages are not shown as they produce relatively weaker lines (for a given U), and are unable to match the observed properties of \gndone\ except for the most extreme case of \bpass binary stellar populations with an IMF that extends up to 300 M$_{\odot}$.  \editone{If we allow for a lower C/O abundance ratio (compared to  our assumed Solar value), then the \wciii would be lower for all models (see discussion of \citealt{jask16} and \citealt{gutk16}, who consider changes in abundance ratios in low metallicity high-redshift sources).}  

As illustrated in Figure~\ref{fig:cloudy-ciii-irac}, stellar population models that lack binaries and very high-mass stars have difficulty simultaneously producing the measured \wciii and IRAC [3.6]--[4.5] color.  Neither the S99 nor the \bpass models without binaries with an IMF that extends to 100 M$_{\odot}$ can reproduce the observed data -- unless the nebular gas has a very low metallicity ($<$0.1 Z$_\odot$), and very high ionization (log\,U $\gtrsim -1.5$), and then only if the IRAC color lies at the lower end of its error distribution.  \bpass models that include the stellar binaries and/or an upper mass of the IMF that extends to M$_{\text{IMF}}^{up} =$ 300 M$_{\odot}$ produce harder spectra and better match the observed \wciii and [3.6]--[4.5] color over a larger range of ionization parameter, log\,U = $-2.1$ to $-$1.5.  These models, which include the effects of stellar binaries, favor lower metallicity for both the ionizing stellar population and the nebular gas, with $Z \simeq 0.1-0.3~ Z_\odot$.   All the models strongly disfavor higher metallicity, $>$0.5 Z$_\odot$, for either the stellar or gas components as these would produce much lower \wciii than observed.    Therefore, the data favor \bpass models with binaries with an IMF that extends to higher-mass stars and low metallicities.  The effects of dust attenuation do not change these conclusions.  \editone{While these were our youngest models, using stellar population models younger than 10 Myrs would also produce higher ionization, resulting in more models that could possibly reach into the space occupied by our galaxy.} 

The results of our models are consistent with other work in the literature.  In a series of simulations using \cloudy and various S99 and \bpass SP models, \citet{jask16} found that \wciii increased with increasing U at all $Z_*$.   \wciii was also found to slightly increase with increasing nebular density (traced by total Hydrogen density), which follows the assumption that \ciii\ traces denser regions \citep[also related to a higher U; e.g.][]{sand16,stei16,jask16,berg18}.   The youngest models in our sample (10 Myrs) fit within \citeauthor{jask16}'s expected range for strong \wciii with a fairly high ionization parameter (log\,U $\geq -2$) and continuous star formation, as well as lending weight to the notion that binary stellar evolution may be necessary to effectively reproduce these observed properties \citep[also suggested by e.g.][]{stei16}.\footnote{\editone{Note that our models assume Solar abundance ratios (specifically, C/O=0.51) and continuous star-formation.  These produce higher \ciii\ equivalent width compared to either models with lower abundance ratios (C/O) or instantaneous burst star-formation, see, for example, discussion in \citet[their Figures 4 and 16]{jask16}.}}   A similar result was found at lower redshifts by \citet{berg18} with a z$\sim$2 lensed galaxy, with a best fit model involving binary stars, log\,U = $-1.5$, $Z=0.05~Z_{\odot}$, and an age of 10 Myrs (however their models used \bpass SPs with an instantaneous burst instead of continuous star formation).

We also considered ionization in \gndone\ from an active galactic nucleus (AGN), as this could boost both the \ciii\ and \oiii\ emission \citep[e.g.,][]{jask16,mase17}.   From the the deep G102 grism data from Faint Infrared Grism Survey \citep[FIGS;][]{pirz17}, T16 find tentative evidence for weak \nv\ emission in \gndone, which could suggest ionization from a weak AGN (however, this \nv\ emission was not detected by the $>$16 hr Keck/MOSFIRE spectrum in J19). We ran AGN models using {\sc Cloudy}'s \texttt{table\,\,agn}\footnote{described by Table 6.3 of Hazy, the \cloudy documentation.} SED, with parameter values and ranges listed in Table \ref{table:2}.

\begin{figure}[t]
\centering
\includegraphics[width=\linewidth]{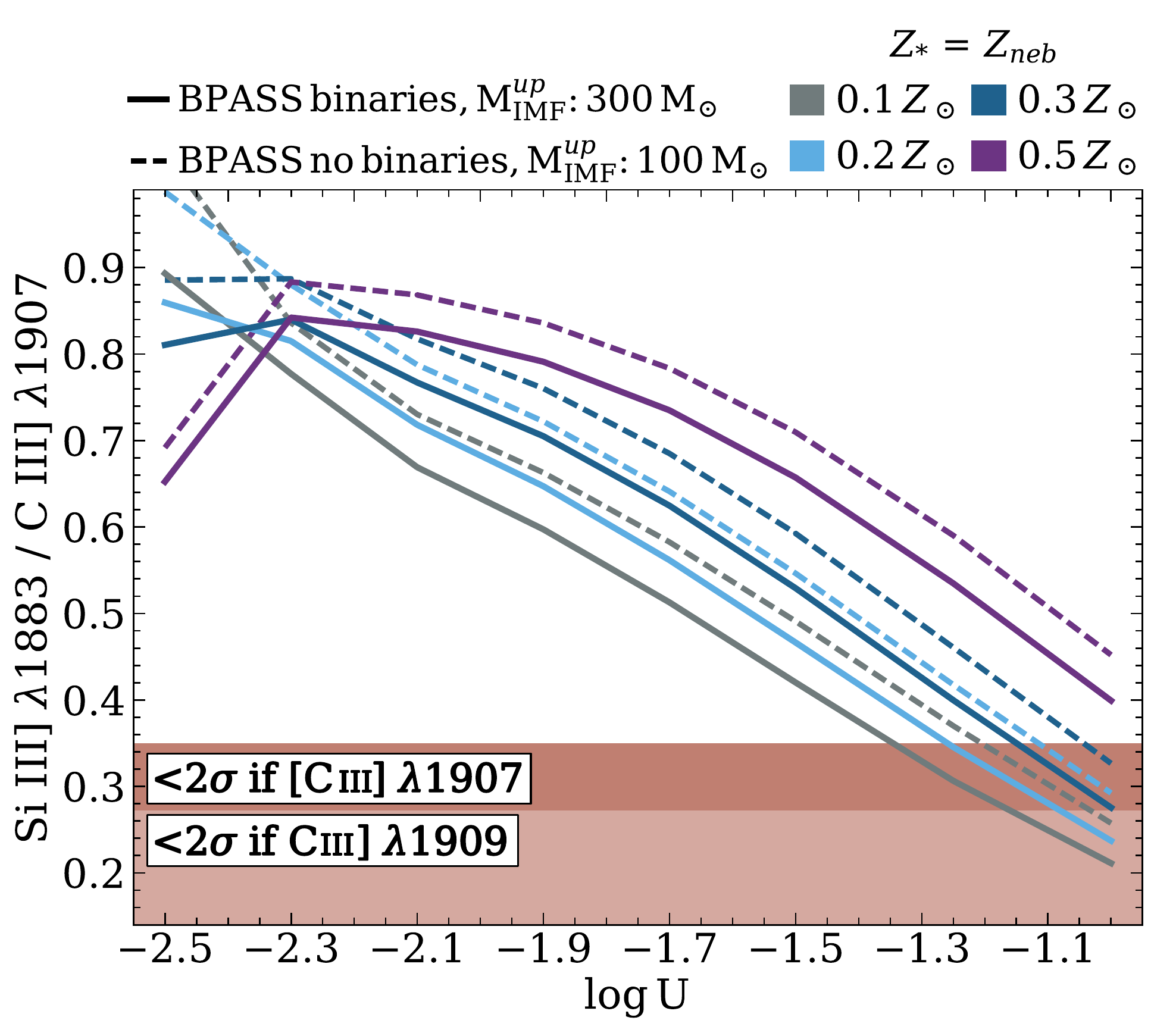}
\caption{The ratio of \siiii\ $\lambda$1883 / [\ciii\ $\lambda$1907 versus ionization parameter for our \editone{fiducial {\sc bpass}+\cloudy models with continuous star formation and age of 10 Myr.  The solid lines show models that include binaries with an IMF that extends to 300 M$_\odot$, and the dashed lines shows models without binaries and an IMF that extends to 100 M$_\odot$.}  The line colors match the stellar metallicities shown in Figure \ref{fig:cloudy-ciii-irac}, and the shaded regions represent the 2$\sigma$ upper limits for this ratio assuming the two different $z_{sys}$ solutions. \label{fig:U-si3c3}}
\end{figure}

None of the AGN models simultaneously produce the \wciii and IRAC color  as measured in \gndone.   The bottom-right panel of Figure~\ref{fig:cloudy-ciii-irac} shows the expected \wciii and IRAC [3.6]--[4.5] color for the \cloudy runs with the AGN spectrum for a range of gas metallicity and ionization parameter.   To reproduce the IRAC color observed in \gndone\ requires higher ionization, which overproduces \wciii for these models.   Some of this is mitigated if the line detected is \ciii\ $\lambda$1909, and the IRAC color is at the low end of its error distribution (or if there is substantially more dust than inferred from the analysis of the SED, see F13).    Therefore, it seems unlikely that the ionization in \gndone\ is powered solely by an AGN.   One possibility is that the ionization in \gndone\ is powered by a composite star-forming stellar population and an AGN.  This has been observed in some galaxies.  For their sample of $z$ = 2--4 \ciii-emitters, \citet{naka18a} found that high \wciii ($\geq$ 20\ang) required a composite model of AGN and SPs to fit this and other galaxy properties.   The presence of an AGN in \gndone\ would have interesting implications for galaxies at $z > 6.5$, as some ionization from AGN may be required for the latter half of reionization \citep[e.g.,][]{finl16,mitra18,fink19}.    The current dataset for \gndone\ prevents exploring more complex models (such as an AGN/stellar composite), but future studies (e.g., with \jwst) will allow this. 

The photoionization models make predictions for other nebular emission lines that could be present, including \siiii$\lambda$1883, for which we place an upper limit on the line flux from the MOSFIRE $H$--band spectrum (see above).    Figure~\ref{fig:U-si3c3} compares our observed \siiii/[\ciii\ limit to the predicted values from our \editone{youngest \cloudy models (10 Myr)}.  
For  Solar abundances, the photoionization models produce a ratio of (\siiii\ $\lambda$1883)/([\ciii\ $\lambda$1907) that varies over a range of 0.4--1.0 (for $-3.5 < {\rm log\,U} < -1.5$), higher than our measured upper limit.  Indeed, the measured limit ($2\sigma$) on the  ratio, \siiii/[\ciii\ $<$0.35,  requires both high ionization (log\,U $\gtrsim -1.5$) and lower metallicity ($Z\simeq 0.1-0.2$~$Z_\odot$).  
We note that the results from the photoionization modeling assume Solar abundance ratios.  Using a lower (sub-Solar) [Si/C] abundance could also reduce the \siiii/[\ciii\ flux ratio \citep[as found by][]{berg18} bringing the models more in line with the data.     Therefore, the \textit{lack} of detected \siiii\ supports the conclusion that the overall ionization in \gndone\ must be extremely high combined with lower metallicity.  This is consistent with the results derived from the analysis of the total \ciii\ equivalent width and IRAC [3.6]--[4.5] color, above.

To summarize, \gndone\ appears to require a hard ionizing stellar population and lower metallicity to reproduce the total \ciii\ equivalent width and IRAC [3.6]--[4.5] color.   The data further suggest that abundances of \gndone\ for elements like Silicon are sub-Solar compared to Carbon and Oxygen, consistent with other findings at $z\sim 2$ \citep[e.g.,][]{berg18}.  In-depth studies must await additional data from facilities such as \jwst.

\begin{figure*}[th]
\centering
\includegraphics[width=0.85\textwidth]{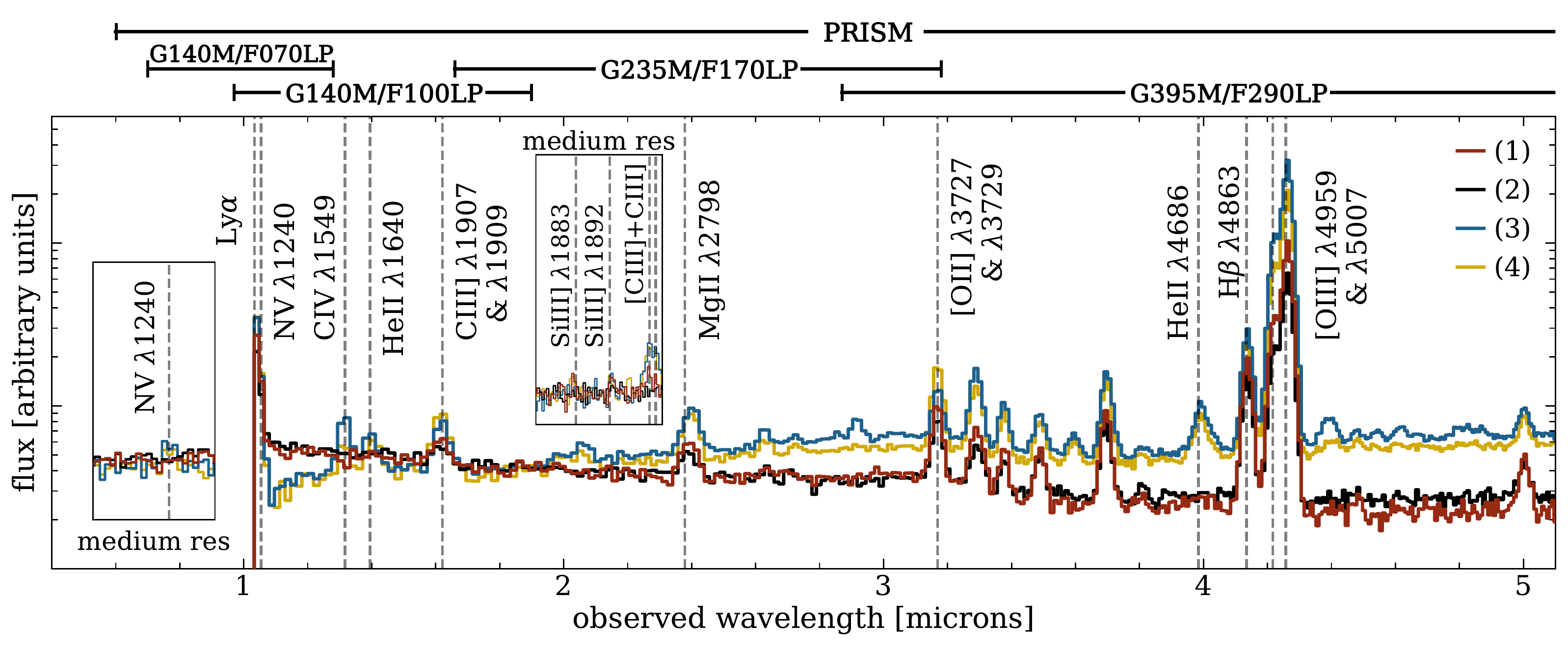}
\caption{Simulated 1D spectra for \gndone\ for a 2 hour (prism) and a 10 hr integration (zoom-in on \nv\ \editone{and \siiii\ $+$ \ciii} in the insets) using NIRSpec on {\it JWST}. (1) Our fiducial SP+\cloudy model with binaries, M$_{\text{IMF}}^{up}=300$ M$_{\odot}$, log\,U $=-2.1$ and $Z_* = Z_{\rm neb} =$ 0.2 $Z_{\odot}$. (2) An SP+\cloudy model with no binaries, M$_{\text{IMF}}^{up}=100$ M$_{\odot}$, and matched log\,U, $Z_*$, and $Z_{\rm neb}$ to our fiducial model. The two AGN models have (3) log\,U $=-1.7$ and $Z_{\rm neb}=0.5$  $Z_{\odot}$, and (4) log\,U $=-2.3$ and $Z_{\rm neb}=0.3$ $Z_{\odot}$.  Major lines of interest are labeled, and the spectral coverage for the prism and medium resolution gratings are shown at the top. \label{fig:jwst-etc}}
\end{figure*}

\begin{deluxetable*}{cccccccccc}[thp]
\tablecaption{\label{table:3}Estimated Line Fluxes Relative to H$\beta$ for {\sc Cloudy}+SP/AGN Models with the {\it JWST}/NIRSpec R$\sim$100 Prism.}
\tablecolumns{10}
\tablehead{
\multirow{2}{*}{model} & \multirow{2}{*}{{\sc C\,iv}$\lambda1548$} & 
\multirow{2}{*}{He{\sc\,ii}$\lambda1640$} & 
\multirow{2}{*}{$\substack{\text{Si{\sc\,iii}]}\lambda1883,1892 +\\\text{C{\sc\,iii}]}\lambda1907,1909}$} & 
\multirow{2}{*}{Mg{\sc\,ii}$\lambda2798$} & 
\multirow{2}{*}{{\sc [O\,ii]}$\lambda3727,3729$} & 
\multirow{2}{*}{He{\sc\,ii}$\lambda4686$} & 
\multirow{2}{*}{H$\beta$} & 
\multirow{2}{*}{{\sc [O\,iii]}$\lambda4959$} & 
\multirow{2}{*}{{\sc [O\,iii]}$\lambda5007$} \\
~\vspace{-6mm}\\}

\startdata
(1) & \dotfill &  0.05$^{+0.03}_{-0.02}$ &  0.30$^{+0.12}_{-0.05}$ &  0.29$^{+0.03}_{-0.05}$ &  0.43$^{+0.02}_{-0.02}$ &  0.15$^{+0.03}_{-0.03}$ &  1.00 &  1.82$^{+0.04}_{-0.04}$ &  5.55$^{+0.08}_{-0.11}$ \\
(2) &  \dotfill &  0.07$^{+0.02}_{-0.03}$ &  0.42$^{+0.11}_{-0.11}$ &  0.26$^{+0.03}_{-0.03}$ &  0.47$^{+0.03}_{-0.02}$ &  0.18$^{+0.04}_{-0.04}$ &  1.00 &  1.56$^{+0.05}_{-0.03}$ &  4.73$^{+0.09}_{-0.12}$ \\
(3) & 0.32$^{+0.03}_{-0.02}$ &  0.22$^{+0.49}_{-0.03}$ &  0.28$^{+0.02}_{-0.02}$ &  0.33$^{+0.02}_{-0.02}$ &  0.30$^{+0.01}_{-0.01}$ &  0.34$^{+0.02}_{-0.02}$ &  1.00 &  4.15$^{+0.05}_{-0.05}$ &  12.48$^{+0.12}_{-0.16}$ \\
(4) & 0.18$^{+0.06}_{-0.04}$ &  0.23$^{+0.05}_{-0.05}$ &  0.45$^{+0.02}_{-0.03}$ &  0.37$^{+0.02}_{-0.02}$ &  0.63$^{+0.02}_{-0.01}$ &  0.30$^{+0.03}_{-0.02}$ &  1.00 &  3.04$^{+0.03}_{-0.04}$ &  9.23$^{+0.1}_{-0.12}$ \\
\enddata
\tablecomments{The line flux ratios (relative to H$\beta$) shown are the median, 16th percentile, and 84th percentile measurements based upon our 2hr {\it JWST}/NIRSpec R$\sim$100 prism simulations. (1) Fiducial SP+\cloudy model with binaries and M$_{\text{IMF}}^{up}$ = 300 M$_\odot$, log\,U $=-2.1$ and $Z_* = Z_{\rm neb} =$ 0.2 $Z_{\odot}$. (2) SP+\cloudy model with no binaries and M$_{\text{IMF}}^{up}$ = 100 M$_\odot$, log\,U $=-2.1$ and $Z_* = Z_{\rm neb} =$ 0.2 $Z_{\odot}$. ~(3) AGN model with log\,U $=-1.7$ and $Z_{\rm neb}=0.5$ $Z_{\odot}$. ~(4) AGN model with log\,U $=-2.3$ and $Z_{\rm neb}=0.3$ $Z_{\odot}$.}
\end{deluxetable*}

\subsection{Predictions for \jwst} \label{subsec:jwst}

The data for \gndone\ favor stellar population models with high-mass stars as in the \bpass models, with an IMF that extends up to 300 M$_{\odot}$, binary stellar populations, and lower metallicities.  This fits well with current expectations that galaxies in the early Universe are younger, more metal-poor systems \citep[e.g.,][]{star15b,stei16,jask16,berg18}.  However, alternative explanations remain.    To test our conclusions, we make predictions for the advanced capabilities of future instruments such as NIRSpec on the {\it James Webb Space Telescope} ({\it JWST}).  The different models make strongly varying predictions for nebular emission expected for this galaxy.    Detecting strong \heii\ and \nv\ emission could provide insight into the nature of the ionizing source of \gndone, allowing a more definitive distinction between the nature of the stellar populations, the presence of an obscured AGN, the combination of the two, or something wholly unexpected.  

We make predictions for {\it JWST}/NIRSpec using our fiducial model (model 1; \bpass models with binaries with an IMF that extends to 300~$M_\odot$, constant SFR for an age of 10 Myr, and $Z_* = 0.2$~$Z_\odot$) that matched the galaxy's ionization conditions as well as three other models.  These include another SP model (model 2) with no binaries, an IMF that extends to 100~$M_\odot$, and matched parameters to our fiducial model (for comparison), and two AGN models from our analysis above with \cloudy parameters: (model 3)  log\,U $=-1.7$, $Z_{\rm neb}=0.5$ $Z_{\odot}$,  and (model 4) log\,U $=-2.3$, $Z_{\rm neb}=0.3$ $Z_{\odot}$.   In both models 1 and 2, we set the SP+\cloudy models to have log\,U $=-2.1$ and $Z_*=Z_{\rm neb}=0.2~Z_\odot$.  We run these models through the {\it JWST} exposure time calculator (ETC)\footnote{\url{https://jwst.etc.stsci.edu}} via the \texttt{Python} Pandeia Engine\footnote{\url{https://jwst-docs.stsci.edu/display/JPP/Pandeia+Quickstart}} using slitted spectroscopy on NIRSpec. The continuum emission in the spectra were redshifted to our measured systemic value, \editone{$z_{sys}$ = 7.5032}, with flux density normalized to match the updated \hst F160W magnitude for \gndone\ ($H_{160} = 25.38$ AB mag, Finkelstein et al.\ in prep). We used a fixed slit setup with the prism and medium resolution gratings for the ETC runs, setting each `target' spectrum in its own scene. The background noise was set medium.

Figure \ref{fig:jwst-etc} shows simulated prism spectra for \gndone\ for NIRSpec with an exposure time of 2 hrs.  Inset panels in the figure showed detail around the wavelengths where we expect \nv\ and \ciii, assuming medium resolution gratings (G140M, G235M, G395M; see top part of Figure \ref{fig:jwst-etc}) for (longer) exposure times of 10 hours. Table~\ref{table:3} provides the predicted line fluxes relative to \hb\ for each model.   Even at the coarse spectroscopic resolution of the prism, it will be \editone{possible to measure} simultaneously \ciii, \oiii, and \lya.  In particular, \hb\ and \oiii\ will be resolved, answering most of the questions about the nature of the ionization as the expected ratio of \oiii/\hb\ is expected to be $\simeq$2$\times$ higher for an AGN (see Table \ref{table:3}).  It may also be possible to determine spatially-varying ionization.  The higher resolution NIRSpec gratings (G140M/G140H, G235M/G235H, and G395M/G395H) should be able to differentiate between these lines with resolving powers of R$\simeq$1,000 and R$\simeq$2,700, respectively. At the higher resolution, more features become distinguishable such as the \hbox{Si\,{\sc iii}]} $\lambda$1883,1893 doublet, where comparisons to lines such as \ciii\ allow for accurate constraints on the gas density and Si/C abundances \citep[e.g.,][]{stei16,berg18}.  

Detections of these lines and any spatial variation in their intensity would further constrain several important properties of this galaxy.  Identification of certain spectral features such as \nv\ and \civ\ would point more explicitly towards a contribution from an AGN.  Resolved measurements of UV diagnostic lines \ciii, \civ, and \heii\ \citep[e.g.,][]{vill97,felt16,naka18a} would independently label the ionizing source as a young actively star-forming galaxy, or an obscured (possibly weakly) accreting AGN, or possibly a composite of the two.

\section{Summary} \label{sec:summary}
Using Keck/MOSFIRE $H$--band spectroscopy of a galaxy at $z_{Ly\alpha}$ = 7.51, previously identified via \lya\ emission by \citet{fink13}, we identified line emission that we associate with one component of the \ciii$\lambda\lambda$1907,1909 doublet at $\lambda_{obs} = 1.6213$~\micron.   We optimally extracted the 1D spectrum and measured a line flux of $(2.63 \pm 0.52) \times 10^{-18}$ erg s$^{-1}$ cm$^{-2}$, with S/N=5.6 integrated over the line.  We do not detect \ion{Si}{3}] $\lambda\lambda$1883, 1892, which yields an upper limit on the line ratio, \ion{Si}{3}]/\ciii $<$ 0.35 ($2\sigma$).

Using the CANDELS photometry and \lya\ measurement, we \editone{tentatively identified the detected line as} the [\ciii\ $\lambda$1907 part of the \ciii\ doublet, providing a systemic redshift, \editone{$z_\mathrm{sys} = 7.5032 \pm 0.0003$}.   This resulted in \vlya\ = \editone{$88\pm27$} km s$^{-1}$, suggesting that \gndone\ potentially has significant hard radiation responsible for ionizing neutral \hi\ gas in the vicinity of this galaxy, which would otherwise shift the \lya\ emission further redwards from $z_{sys}$. The low \vlya\ instead implies that the \lya\ emission remains fairly unaffected by \hi\ until it encounters the IGM, suppressing the strength of the emission. In addition the hard radiation could be boosting \ciii\ emission, making the ([\ciii+\ciii)/\lya\ ratio one of the highest measured for high redshift sources.

To investigate the properties of \gndone\ further, we modeled the expected total \ciii\ equivalent width and IRAC [3.6]--[4.5] color using models (where the IRAC color is a measure of the redshifted \hb\ + \oiii\ emission).  We explored both \bpass and S99 stellar population models, and AGN power-law models, using the \cloudy radiative transfer code to predict the nebular emission. We used a range of input stellar models, both single and binary stellar populations, with a range of stellar and nebular metallicities, stellar population age, and ionization parameter.    

The models that best reproduce the observed \wciii and IRAC [3.6]--[4.5] color require stellar populations with models like in {\sc bpass}, that include both the contributions from binary stars and an IMF that extends up to M$_{\text{IMF}}^{up} =$ 300 M$_{\odot}$,  low stellar metallicity $Z_*$ = 0.1--0.2 $Z_{\odot}$, low nebular metallicity $Z_{\rm neb}$ = 0.2--0.3 $Z_{\odot}$, and high ionization parameter log\,U $\gtrsim -2$.  The upper limit on \siiii/\ciii\ requires even higher ionization, log\,U $\gtrsim -1.5$,  combined with lower metallicty, $Z \simeq 0.1-0.2$~$Z_\odot$.   The pure AGN models do not reproduce simultaneously the \wciii and [3.6]--[4.5] color, but we can not rule out a possible a composite model including both a stellar population plus an AGN component.   

The nature of the ionizing source(s) of \gndone\ will be differentiable with forthcoming spectroscopy with {\it JWST}.  To make predictions for such data, we simulated {\it JWST} spectra for this galaxy using the closest models for \gndone\ for both the case of ionization by stellar populations or AGN and an additional comparison model.  Even the coarse spectroscopy ($R\sim 100$) will resolve \hb\ and \oiii, where we expect large differences in the flux ratio depending on the nature of the ionization.   Other observations at higher resolution will also be able to \editone{separate close doublets (such as \ciii) and} constrain better the metallicity, nature of the ionization, and its spatial variation, opening a new window into the nature of galaxies in the Epoch of Reionization. \\

\acknowledgments
The authors wish to recognize and acknowledge the very significant cultural role and reverence that the summit of Mauna Kea has always had within the indigenous Hawaiian community. We are most fortunate to have the opportunity to conduct observations from this sacred mountain.  The authors thank R. Kennicutt, D. Stark, \editone{A. Jaskot, S. Ravindranath,} D. Berg, and J. Cohn 
for insightful conversations and for constructive comments on the draft. \editone{We thank the anonymous referee for comments that improved this paper.}
TAH and CJP acknowledge generous support from the Texas A\&M University and the George P. and Cynthia Woods Institute for Fundamental Physics and Astronomy. SLF acknowledges support from NSF AAG award AST-1518183. IJ acknowledges support from the NASA Headquarters under the NASA Earth and Space Science Fellowship Program - Grant 80NSSC17K0532. RSE acknowledges funding from the European Research Council (ERC) under the European Union's Horizon 2020 research and innovation programme (grant agreement No 669253). SM and JER acknowledge support from NSF AAG award 1518057.

This work was supported by a NASA Keck PI Data Award, administered by the NASA Exoplanet Science Institute. Data presented herein were obtained at the W. M. Keck Observatory from telescope time allocated to the National Aeronautics and Space Administration through the agency's scientific partnership with the California Institute of Technology and the University of California. The Observatory was made possible by the generous financial support of the W. M. Keck Foundation.

\appendix
The following text concerns the data analysis used in this work. Due to the low signal-to-noise of these data, we believe this is important to share. This section is separated into two subsections: \ref{appsub:phot} discusses how we measured the photometric variability of our data for each observational epoch and \ref{appsub:error} discusses our method for calculating the optimized error for each dataset.\\

\section{Photometric Variability} \label{appsub:phot}
\begin{figure}[!ht]
    \centering
    \begin{minipage}{0.495\textwidth}
        \centering
        \includegraphics[width=\textwidth]{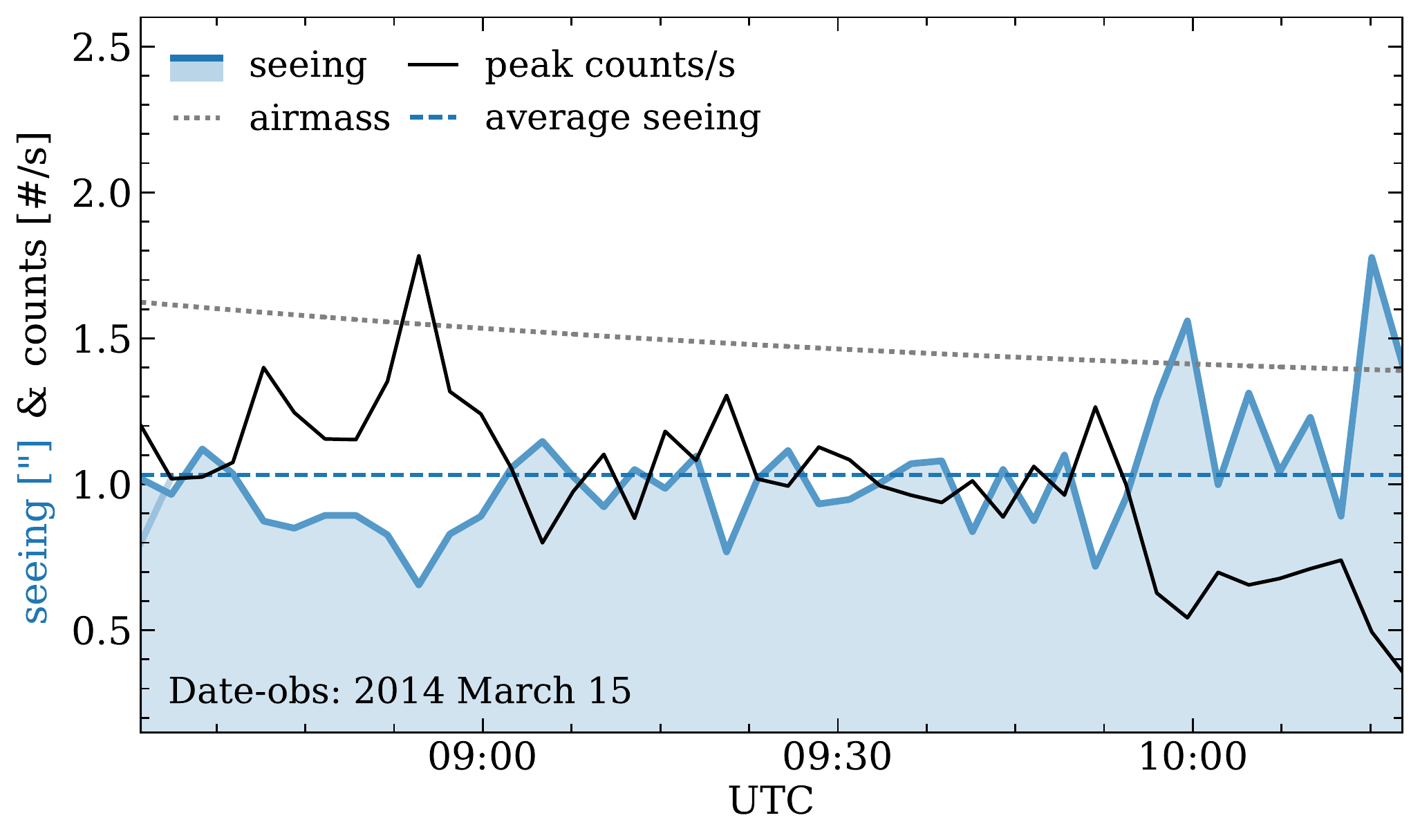}
    \end{minipage}
    \begin{minipage}{0.495\textwidth}
        \centering
        \includegraphics[width=\textwidth]{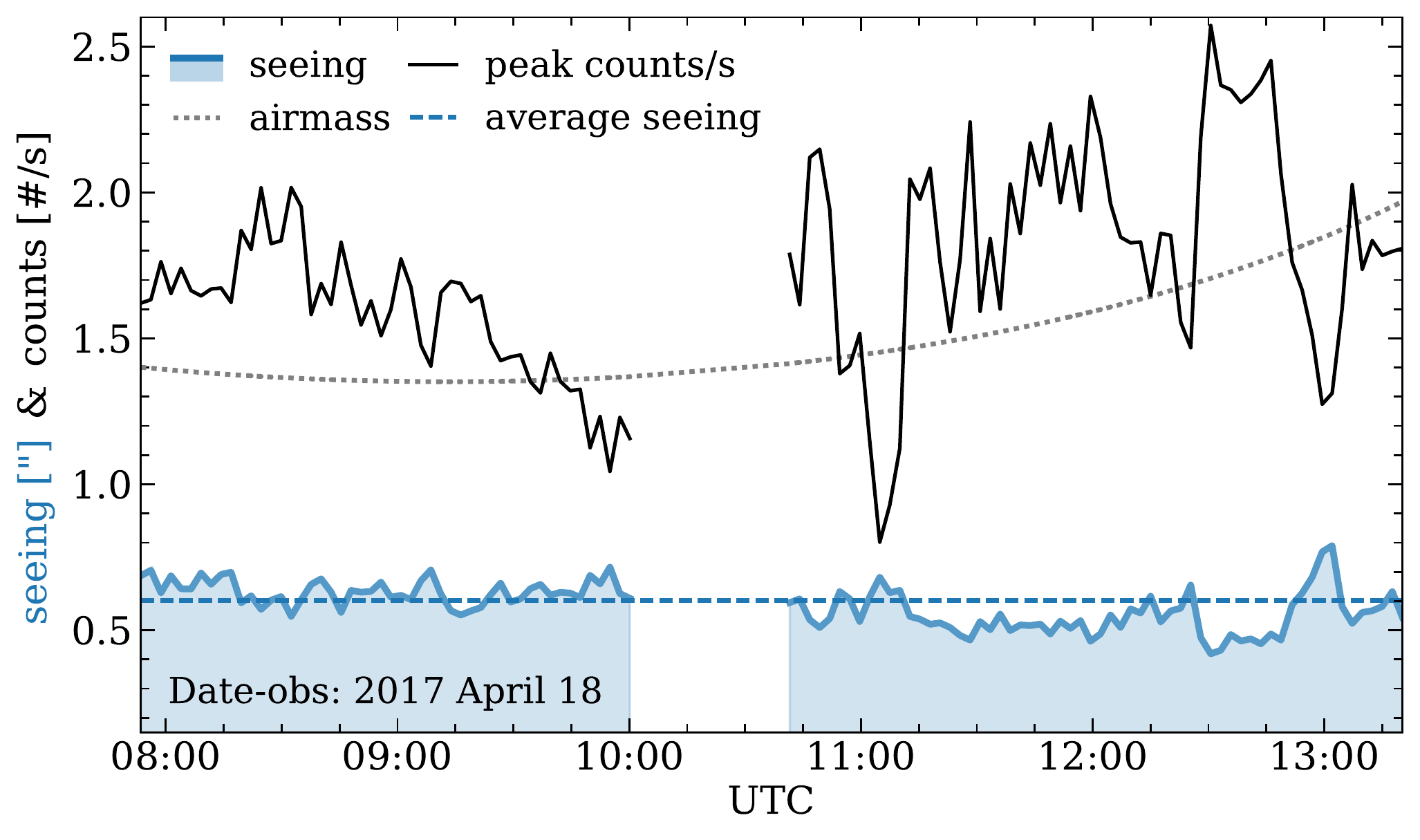}
    \end{minipage}
    \caption{Peak counts and seeing (measured from FWHM of the spatial profile of the reference star) for each raw data frame for 2014 March 15 and 2017 April 18. Overplotted are the airmass and average seeing (measured from the final reduced 2D spectra of the reference star). Due to the significant variability in peak counts and the poor seeing, the 2014 March 15 data were left out of the analysis. \label{fig:seeing}}
\end{figure}

The photometric variability for each epoch was determined by measuring the peak counts and the FWHM of the spatial profile of the `reference star' in the mask in each raw 120--180s frame. To measure the values, we chose a region along the raw 2D spectra devoid of skylines to sum over in order to increase the signal -- the same region was used for all 2014 data, while a similar region was chosen for the 2017 data. Figure \ref{fig:seeing} shows the results of the 2014 March 15 variability (left), with the 2017 April 18 data (right) shown for comparison. Overplotted on both are the measured airmass over the full exposure and the average seeing, measured from the FWHM of the spatial profile of the final reduced 2D spectra for the reference star from each mask.

On a night with fair to good seeing, one would expect the peak counts and seeing to remain relatively constant; on a night with significant clouds or other atmospheric effects, the peak counts may vary wildly, with the seeing either remaining fairly constant or varying inversely to the peak counts. As can be seen in the left panel of Figure \ref{fig:seeing}, the 2014 March 15 data showed significant variability in peak counts per raw frame -- with persistent poor seeing averaging to 1\farcs03, growing worse as the target field approached zenith. For comparison, the 2017 April 18 data showed some variability in peak counts, but not at all to the degree shown for the former dataset; in addition, the seeing remained fairly constant, averaging to 0\farcs602. Due to this evidence, we determined that the 2014 March 15 data should be left out of the analysis so as to not add unnecessary noise to our measurements.

\begin{figure}[!ht]
    \centering
    \begin{minipage}{0.495\textwidth}
        \centering
        \includegraphics[width=\textwidth]{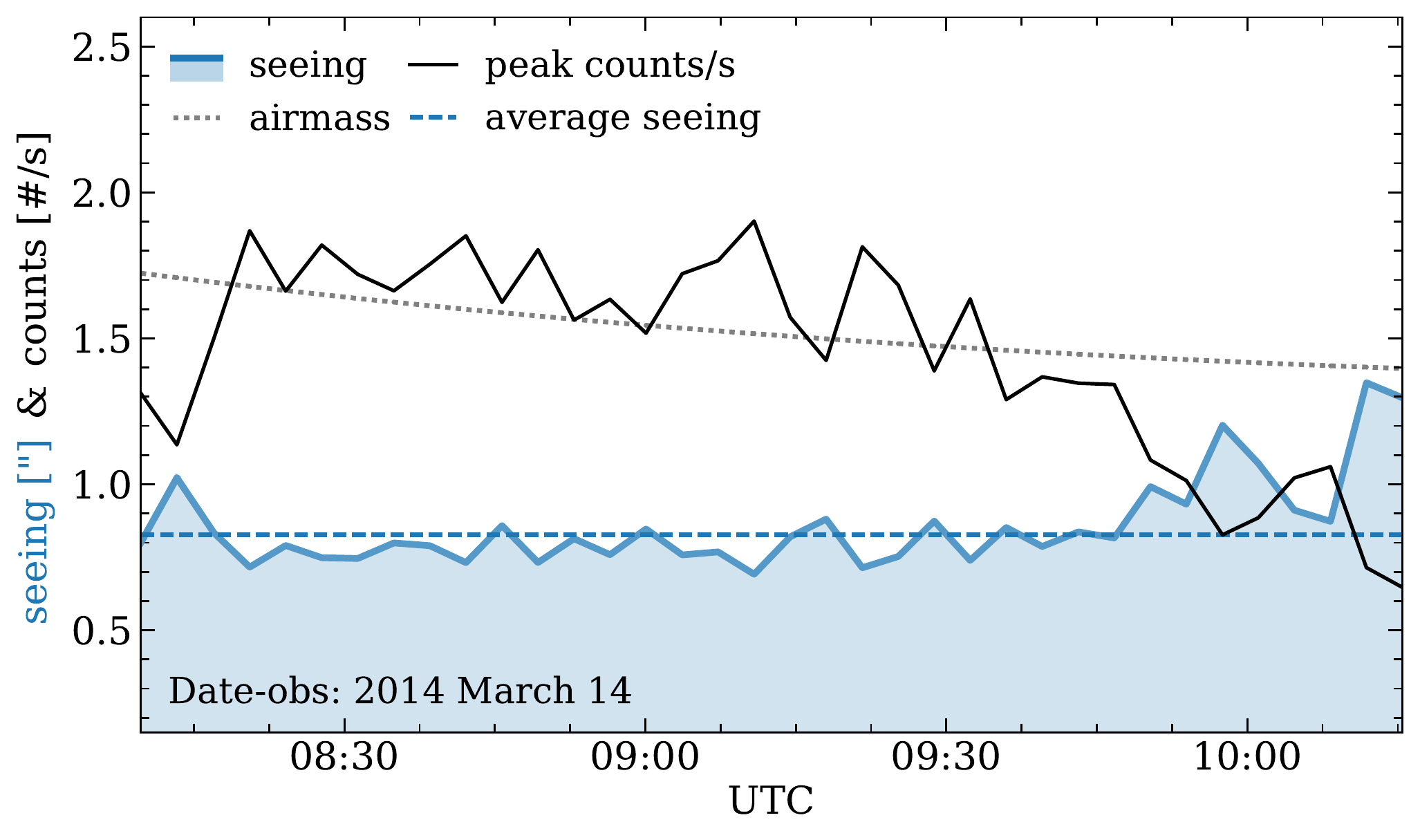}
    \end{minipage}
    \begin{minipage}{0.495\textwidth}
        \centering
        \includegraphics[width=\textwidth]{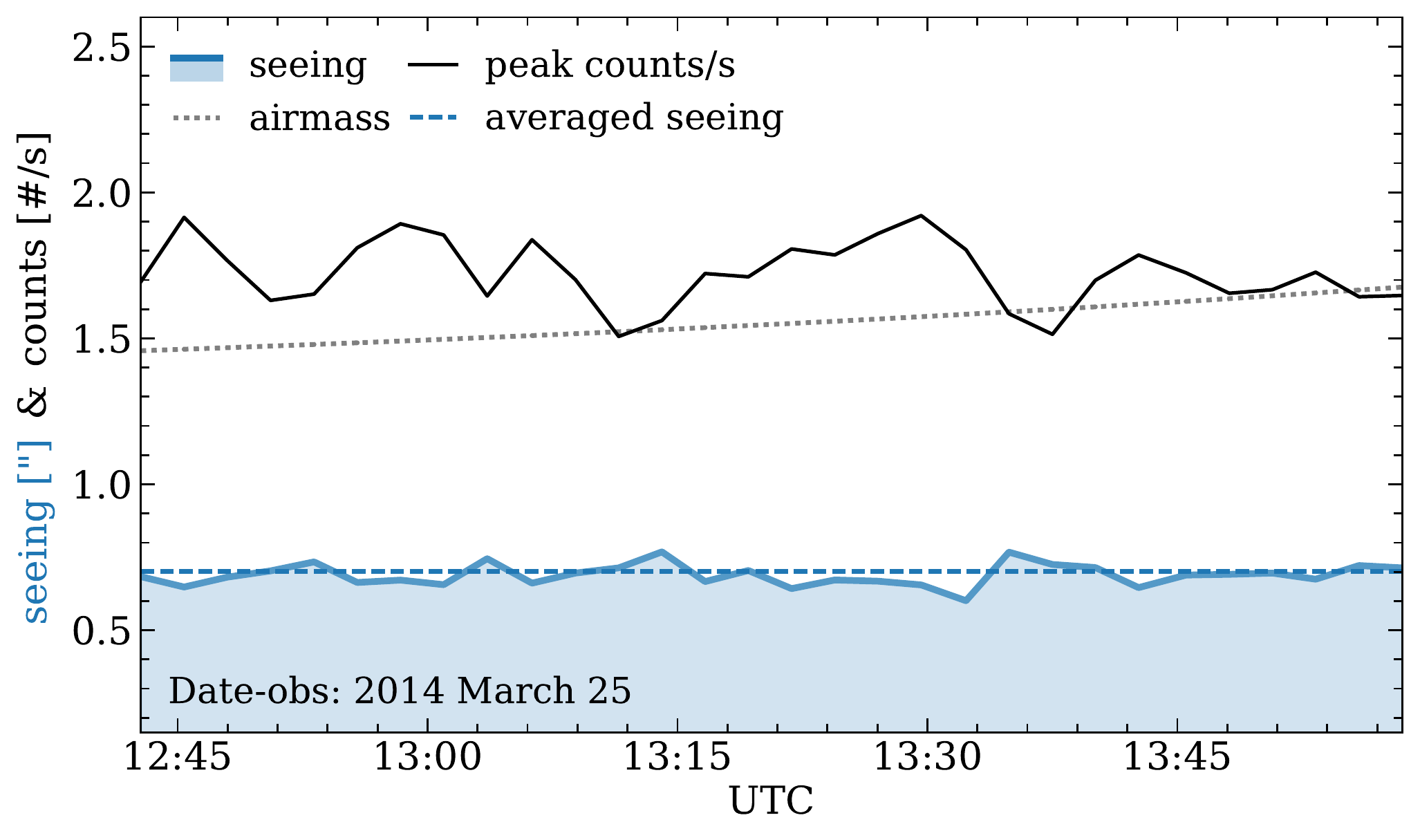}
    \end{minipage}
    \caption{Peak counts and seeing (measured from FWHM of the spatial profile of the reference star) for each raw data frame for 2014 March 14 \& 25. Overplotted are the airmass and average seeing (measured from the final reduced 2D spectra of the reference star). Due to the significant variability in peak counts from the median, only 08:20--09:30 UTC from the 2014 March 14 data were used in the analysis. \label{fig:seeing14}}
\end{figure}

In addition to that epoch, a handful of frames were removed from the 2014 March 14 data where the peak counts varied significantly from the median. Figure \ref{fig:seeing14} shows these fluctuations in the 2014 March 14 data (left), as well as the pristine 2014 March 25 data (right). Due to these results, we only used 08:20--09:30 UTC from the 2014 March 14 data.

\section{Optimized 1D Error} \label{appsub:error}
For the optimized 1D error associated with the optimized 1D spectra, we used an alternative approach than described by Horne (1986). Due to the nature of the multi-slit mask of targets created by MOSFIRE, we calculated an error that would be applicable for every object in the mask. Firstly, as this was primarily a search for high-redshift galaxies, a few of our targets yielded non-detections, providing full slit space to utilize for our optimized error.

\begin{figure*}[th]
\centering
\includegraphics[width=\textwidth]{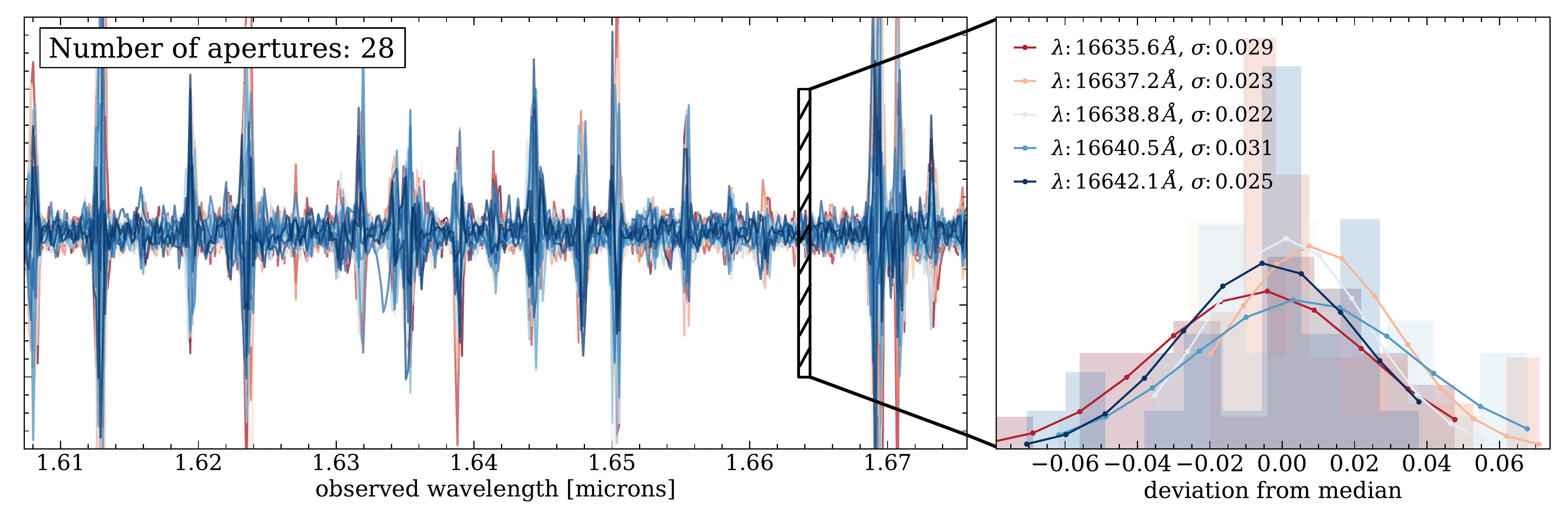}
\caption{{\it Left:} A zoom-in of a section of the spectral range for the `blank' spectra for the 2017 April 18 data, showing noise and skylines. The hatched rectangle indicates the spectral range shown on the right. {\it Right:} Section of `blank' spectra showing the distribution of the apertures and the standard deviation from the median value of counts per wavelength step. \label{fig:histerror}}
\end{figure*}

Following the same optimized 1D extraction technique developed and used on our target spectra, we optimally extracted `blank' apertures in these regions in the full mask devoid of any discernible signal or negative trace (unique to spectrographs like MOSFIRE). Building a statistical sample of these `blank' apertures, we then plotted their distribution at each wavelength step and used the resulting standard deviation as the error at that wavelength step. The left panel of Figure \ref{fig:histerror} shows all of the `blank' apertures for the 2017 April 18 data -- as expected the resulting spectra only shows the skylines and noise of the data. The right panel of Figure \ref{fig:histerror} shows an example of the distributions at different wavelength steps and their resulting statistics.

\begin{thebibliography}{}

\bibitem[Ba{\~n}ados et al.(2018)]{bana18} Ba{\~n}ados, E., Venemans, B.~P., Mazzucchelli, C., et al.\ 2018, \nat, 553, 473


\bibitem[Berg et al.(2018)]{berg18} Berg, D.~A., Erb, D.~K., Auger, M.~W., Pettini, M., \& Brammer, G.~B.\ 2018, \apj, 859, 164


\bibitem[Bouwens et al.(2015)]{bouw15} Bouwens, R.~J., Illingworth, G.~D., Oesch, P.~A., et al.\ 2015, \apj, 803, 34


\bibitem[Byler et al.(2018)]{byler18} Byler, N., Dalcanton, J.~J., Conroy, C., et al.\ 2018, \apj, 863, 14 


\bibitem[Byler et al.(2017)]{byler17} Byler, N., Dalcanton, J.~J., Conroy, C., \& Johnson, B.~D.\ 2017, \apj, 840, 44 


\bibitem[Calzetti et al.(2000)]{calz00} Calzetti, D., Armus, L., Bohlin, R.~C., et al.\ 2000, \apj, 533, 682


\bibitem[Chonis et al.(2013)]{chon13} Chonis, T.~S., Blanc, G.~A., Hill, G.~J., et al.\ 2013, \apj, 775, 99


\bibitem[Choudhury et al.(2015)]{chou15} Choudhury, T.~R., Puchwein, E., Haehnelt, M.~G., \& Bolton, J.~S.\ 2015, \mnras, 452, 261 


\bibitem[Dijkstra(2017)]{dijk17} Dijkstra, M.\ 2017, arXiv:1704.03416 


\bibitem[Dijkstra(2014)]{dijk14} Dijkstra, M.\ 2014, \pasa, 31, e040 


\bibitem[Ding et al.(2017)]{ding17} Ding, J., Cai, Z., Fan, X., et al.\ 2017, \apjl, 838, L22


\bibitem[Du et al.(2018)]{du18} Du, X., Shapley, A.~E., Reddy, N.~A., et al.\ 2018, \apj, 860, 75


\bibitem[Elbaz et al.(2017)]{elbaz17} Elbaz, D., Leiton, R., Nagar, N., et al.\ 2017, arXiv:1711.10047 


\bibitem[Eldridge et al.(2017)]{eldr17} Eldridge, J.~J., Stanway, E.~R., Xiao, L., et al.\ 2017, \pasa, 34, e058 


\bibitem[Erb et al.(2010)]{erb10} Erb, D.~K., Pettini, M., Shapley, A.~E., et al.\ 2010, \apj, 719, 1168


\bibitem[Erb et al.(2014)]{erb14} Erb, D.~K., Steidel, C.~C., Trainor, R.~F., et al.\ 2014, \apj, 795, 33 


\bibitem[Fan et al.(2002)]{fan02} Fan, X., Narayanan, V.~K., Strauss, M.~A., et al.\ 2002, \aj, 123, 1247 


\bibitem[Feltre et al.(2016)]{felt16} Feltre, A., Charlot, S., \& Gutkin, J.\ 2016, \mnras, 456, 3354


\bibitem[Ferland et al.(2017)]{ferl17} Ferland, G.~J., Chatzikos, M., Guzm{\'a}n, F., et al.\ 2017, \rmxaa, 53, 385 


\bibitem[Finkelstein et al.(2013)]{fink13} Finkelstein, S.~L., Papovich, C., Dickinson, M., et al.\ 2013, \nat, 502, 524 


\bibitem[Finkelstein(2016)]{fink16} Finkelstein, S.~L.\ 2016, \pasa, 33, e037 




\bibitem[Finkelstein et al.(2015)]{fink15} Finkelstein, S.~L., Song, M., Behroozi, P., et al.\ 2015, \apj, 814, 95 


\bibitem[Finkelstein et al.(2019)]{fink19} Finkelstein, S.~L., D'Aloisio, A., Paardekooper, J.~P., Ryan Jr, R., et al.\ 2019, \apj\ Submitted


\bibitem[Finlator et al.(2016)]{finl16} Finlator, K., Oppenheimer, B.~D., Dav{\'e}, R., et al.\ 2016, \mnras, 459, 2299


\bibitem[Giavalisco et al.(2004)]{giav04} Giavalisco, M., Ferguson, H.~C., Koekemoer, A.~M., et al.\ 2004, \apjl, 600, L93 


\bibitem[Grogin et al.(2011)]{grog11} Grogin, N.~A., Kocevski, D.~D., Faber, S.~M., et al.\ 2011, \apjs, 197, 35 


\bibitem[Gutkin et al.(2016)]{gutk16} Gutkin, J., Charlot, S., \& Bruzual, G.\ 2016, \mnras, 462, 1757


\bibitem[Hainline et al.(2011)]{hain11} Hainline, K.~N., Shapley, A.~E., Greene, J.~E., \& Steidel, C.~C.\ 2011, \apj, 733, 31


\bibitem[Hayes et al.(2010)]{hayes10} Hayes, M., {\"O}stlin, G., Schaerer, D., et al.\ 2010, \nat, 464, 562 


\bibitem[Hayes et al.(2011)]{hayes11} Hayes, M., Schaerer, D., {\"O}stlin, G., et al.\ 2011, \apj, 730, 8 


\bibitem[Hayes(2015)]{hayes15} Hayes, M.\ 2015, \pasa, 32, e027 


\bibitem[Hill et al.(2017)]{hill17} Hill, A.~R., Muzzin, A., Franx, M., \& Marchesini, D.\ 2017, \apjl, 849, L26 


\bibitem[Horne(1986)]{horne86} Horne, K.\ 1986, \pasp, 98, 609 


\bibitem[Inoue(2011)]{inoue11} Inoue, A.~K.\ 2011, \mnras, 415, 2920 


\bibitem[Jaskot \& Ravindranath(2016)]{jask16} Jaskot, A.~E., \& Ravindranath, S.\ 2016, \apj, 833, 136 


\bibitem[Jung et al.(2019)]{jung19} Jung, I., Finkelstein, S.~L., Dickinson, M., et al.\ 2019, arXiv:1901.05967


\bibitem[Kimm et al.(2019)]{kimm19} Kimm, T., Blaizot, J., Garel, T., et al.\ 2019, arXiv:1901.05990


\bibitem[Koekemoer et al.(2011)]{koek11} Koekemoer, A.~M., Faber, S.~M., Ferguson, H.~C., et al.\ 2011, \apjs, 197, 36 


\bibitem[Kriek et al.(2015)]{kriek15} Kriek, M., Shapley, A.~E., Reddy, N.~A., et al.\ 2015, \apjs, 218, 15 


\bibitem[Kurucz(1993)]{kuru93} Kurucz, R.\ 1993, SYNTHE Spectrum Synthesis Programs and Line Data.~Kurucz CD-ROM No.~18.~Cambridge, 
Mass.: Smithsonian Astrophysical Observatory, 1993., 18


\bibitem[Laporte et al.(2017a)]{lapo17a} Laporte, N., Ellis, R.~S., Boone, F., et al.\ 2017, \apjl, 837, L21 


\bibitem[Laporte et al.(2017b)]{lapo17b} Laporte, N., Nakajima, K., Ellis, R.~S., et al.\ 2017, \apj, 851, 40 


\bibitem[Larson et al.(2018)]{lars18} Larson, R.~L., Finkelstein, S.~L., Pirzkal, N., et al.\ 2018, \apj, 858, 94


\bibitem[Leitherer et al.(2014)]{leit14} Leitherer, C., Ekstr{\"o}m, S., Meynet, G., et al.\ 2014, \apjs, 212, 14 


\bibitem[Mainali et al.(2017)]{main17} Mainali, R., Kollmeier, J.~A., Stark, D.~P., et al.\ 2017, \apjl, 836, L14


\bibitem[Maseda et al.(2017)]{mase17} Maseda, M.~V., Brinchmann, J., Franx, M., et al.\ 2017, \aap, 608, A4 


\bibitem[Mason et al.(2018)]{mason18} Mason, C.~A., Treu, T., Dijkstra, M., et al.\ 2018, \apj, 856, 2 


\bibitem[Masters et al.(2014)]{mast14} Masters, D., McCarthy, P., Siana, B., et al.\ 2014, \apj, 785, 153 


\bibitem[Matthee et al.(2016)]{matt16} Matthee, J., Sobral, D., Oteo, I., et al.\ 2016, \mnras, 458, 449


\bibitem[Matthee et al.(2017)]{matt17} Matthee, J., Sobral, D., Darvish, B., et al.\ 2017, \mnras, 472, 772 


\bibitem[McGreer et al.(2015)]{mcgr15} McGreer, I.~D., Mesinger, A., \& D'Odorico, V.\ 2015, \mnras, 447, 499 


\bibitem[McLean et al.(2012)]{mcle12} McLean, I.~S., Steidel, C.~C., Epps, H.~W., et al.\ 2012, \procspie, 8446, 84460J 


\bibitem[McLinden et al.(2011)]{mcli11} McLinden, E.~M., Finkelstein, S.~L., Rhoads, J.~E., et al.\ 2011, \apj, 730, 136


\bibitem[McLinden et al.(2014)]{mcli14} McLinden, E.~M., Rhoads, J.~E., Malhotra, S., et al.\ 2014, \mnras, 439, 446


\bibitem[Mitra et al.(2018)]{mitra18} Mitra, S., Choudhury, T.~R., \& Ferrara, A.\ 2018, \mnras, 473, 1416


\bibitem[M{\o}ller \& Warren(1998)]{moll98} M{\o}ller, P., \& Warren, S.~J.\ 1998, \mnras, 299, 661 


\bibitem[Mortlock et al.(2011)]{mort11} Mortlock, D.~J., Warren, S.~J., Venemans, B.~P., et al.\ 2011, \nat, 474, 616 


\bibitem[Nakajima et al.(2018b)]{naka18b} Nakajima, K., Fletcher, T., Ellis, R.~S., Robertson, B.~E., \& Iwata, I.\ 2018, \mnras, 477, 2098


\bibitem[Nakajima et al.(2018a)]{naka18a} Nakajima, K., Schaerer, D., Le F{\`e}vre, O., et al.\ 2018, \aap, 612, A94 


\bibitem[Oesch et al.(2015)]{oesch15} Oesch, P.~A., van Dokkum, P.~G., Illingworth, G.~D., et al.\ 2015, \apjl, 804, L30


\bibitem[Oke \& Gunn(1983)]{oke83} Oke, J.~B., \& Gunn, J.~E.\ 1983, \apj, 266, 713


\bibitem[Ono et al.(2012)]{ono12} Ono, Y., Ouchi, M., Mobasher, B., et al.\ 2012, American Astronomical Society Meeting Abstracts \#220, 220, 429.03 


\bibitem[Papovich et al.(2001)]{papo01} Papovich, C., Dickinson, M., \& Ferguson, H.~C.\ 2001, \apj, 559, 620 


\bibitem[Patr{\'{\i}}cio et al.(2016)]{patr16} Patr{\'{\i}}cio, V., Richard, J., Verhamme, A., et al.\ 2016, \mnras, 456, 4191


\bibitem[Pirzkal et al.(2017)]{pirz17} Pirzkal, N., Malhotra, S., Ryan, R.~E., et al.\ 2017, \apj, 846, 84 


\bibitem[Planck Collaboration et al.(2016)]{plan16} Planck Collaboration, Ade, P.~A.~R., Aghanim, N., et al.\ 2016, \aap, 594, A13 


\bibitem[Raiter et al.(2010)]{rait10} Raiter, A., Schaerer, D., \& Fosbury, R.~A.~E.\ 2010, \aap, 523, A64 


\bibitem[Riess et al.(2016)]{riess16} Riess, A.~G., Macri, L.~M., Hoffmann, S.~L., et al.\ 2016, \apj, 826, 56 


\bibitem[Roberts-Borsani et al.(2016)]{robo16} Roberts-Borsani, G.~W., Bouwens, R.~J., Oesch, P.~A., et al.\ 2016, \apj, 823, 143


\bibitem[Robertson et al.(2015)]{robe15} Robertson, B.~E., Ellis, R.~S., Furlanetto, S.~R., \& Dunlop, J.~S.\ 2015, \apjl, 802, L19 




\bibitem[Sanders et al.(2016)]{sand16} Sanders, R.~L., Shapley, A.~E., Kriek, M., et al.\ 2016, \apj, 816, 23 


\bibitem[Senchyna et al.(2017)]{senc17} Senchyna, P., Stark, D.~P., Vidal-Garc{\'{\i}}a, A., et al.\ 2017, \mnras, 472, 2608


\bibitem[Schaerer(2003)]{scha03} Schaerer, D.\ 2003, \aap, 397, 527 


\bibitem[Schaerer(2002)]{scha02} Schaerer, D.\ 2002, \aap, 382, 28 


\bibitem[Schenker et al.(2012)]{schen12} Schenker, M.~A., Stark, D.~P., Ellis, R.~S., et al.\ 2012, \apj, 744, 179


\bibitem[Schenker et al.(2013)]{schen13} Schenker, M.~A., Ellis, R.~S., Konidaris, N.~P., \& Stark, D.~P.\ 2013, \apj, 777, 67 


\bibitem[Schenker et al.(2014)]{schen14} Schenker, M.~A., Ellis, R.~S., Konidaris, N.~P., \& Stark, D.~P.\ 2014, \apj, 795, 20


\bibitem[Shapley et al.(2003)]{shap03} Shapley, A.~E., Steidel, C.~C., Pettini, M., \& Adelberger, K.~L.\ 2003, \apj, 588, 65


\bibitem[Shibuya et al.(2012)]{shib12} Shibuya, T., Kashikawa, N., Ota, K., et al.\ 2012, \apj, 752, 114


\bibitem[Smit et al.(2015)]{smit15} Smit, R., Bouwens, R.~J., Franx, M., et al.\ 2015, \apj, 801, 122


\bibitem[Sobral et al.(2017)]{sobr17} Sobral, D., Matthee, J., Best, P., et al.\ 2017, \mnras, 466, 1242


\bibitem[Sobral et al.(2018)]{sobr18} Sobral, D., Matthee, J., Darvish, B., et al.\ 2018, \mnras, 477, 2817 


\bibitem[Song et al.(2014)]{song14} Song, M., Finkelstein, S.~L., Gebhardt, K., et al.\ 2014, \apj, 791, 3


\bibitem[Song et al.(2016)]{song16} Song, M., Finkelstein, S.~L., Ashby, M.~L.~N., et al.\ 2016, \apj, 825, 5 


\bibitem[Speagle et al.(2014)]{spea14} Speagle, J.~S., Steinhardt, C.~L., Capak, P.~L., \& Silverman, J.~D.\ 2014, \apjs, 214, 15 


\bibitem[Stanway et al.(2016)]{stan16} Stanway, E.~R., Eldridge, J.~J., \& Becker, G.~D.\ 2016, \mnras, 456, 485 


\bibitem[Stanway et al.(2014)]{stan14} Stanway, E.~R., Eldridge, J.~J., Greis, S.~M.~L., et al.\ 2014, \mnras, 444, 3466 


\bibitem[Stark et al.(2014)]{star14} Stark, D.~P., Richard, J., Siana, B., et al.\ 2014, \mnras, 445, 3200


\bibitem[Stark(2016)]{star16} Stark, D.~P.\ 2016, \araa, 54, 761 


\bibitem[Stark et al.(2017)]{star17} Stark, D.~P., Ellis, R.~S., Charlot, S., et al.\ 2017, \mnras, 464, 469 


\bibitem[Stark et al.(2015a)]{star15a} Stark, D.~P., Richard, J., Charlot, S., et al.\ 2015, \mnras, 450, 1846 


\bibitem[Stark et al.(2015b)]{star15b} Stark, D.~P., Walth, G., Charlot, S., et al.\ 2015, \mnras, 454, 1393 


\bibitem[Steidel et al.(2011)]{stei11} Steidel, C.~C., Bogosavljevi{\'c}, M., Shapley, A.~E., et al.\ 2011, \apj, 736, 160 


\bibitem[Steidel et al.(2016)]{stei16} Steidel, C.~C., Strom, A.~L., Pettini, M., et al.\ 2016, \apj, 826, 159 


\bibitem[Strom et al.(2017)]{strom17} Strom, A.~L., Steidel, C.~C., Rudie, G.~C., et al.\ 2017, \apj, 836, 164


\bibitem[Tang et al.(2018)]{tang18} Tang, M., Stark, D., Chevallard, J., \& Charlot, S.\ 2018, arXiv:1809.09637


\bibitem[Tilvi et al.(2014)]{tilvi14} Tilvi, V., Papovich, C., Finkelstein, S.~L., et al.\ 2014, \apj, 794, 5


\bibitem[Tilvi et al.(2016)]{tilvi16} Tilvi, V., Pirzkal, N., Malhotra, S., et al.\ 2016, \apjl, 827, L14 


\bibitem[Treu et al.(2013)]{treu13} Treu, T., Schmidt, K.~B., Trenti, M., Bradley, L.~D., \& Stiavelli, M.\ 2013, \apjl, 775, L29


\bibitem[Tumlinson \& Shull(2000)]{tuml00} Tumlinson, J., \& Shull, J.~M.\ 2000, \apjl, 528, L65 


\bibitem[Vanzella et al.(2011)]{vanz11} Vanzella, E., Pentericci, L., Fontana, A., et al.\ 2011, \apjl, 730, L35


\bibitem[Venemans et al.(2013)]{vene13} Venemans, B.~P., Findlay, J.~R., Sutherland, W.~J., et al.\ 2013, \apj, 779, 24 



\bibitem[Verhamme et al.(2006)]{verh06} Verhamme, A., Schaerer, D., \& Maselli, A.\ 2006, \aap, 460, 397


\bibitem[Verhamme et al.(2015)]{verh15} Verhamme, A., Orlitov{\'a}, I., Schaerer, D., \& Hayes, M.\ 2015, \aap, 578, A7 


\bibitem[Villar-Martin et al.(1997)]{vill97} Villar-Martin, M., Tadhunter, C., \& Clark, N.\ 1997, \aap, 323, 21


\bibitem[Willott et al.(2015)]{will15} Willott, C.~J., Carilli, C.~L., Wagg, J., \& Wang, R.\ 2015, \apj, 807, 180


\bibitem[Yang et al.(2017)]{yang17} Yang, H., Malhotra, S., Gronke, M., et al.\ 2017, \apj, 844, 171


\bibitem[Zitrin et al.(2015)]{zitr15} Zitrin, A., Labb{\'e}, I., Belli, S., et al.\ 2015, \apjl, 810, L12


\end{thebibliography}
\end{document}